\documentclass[onecolumn]{IEEEtran}

\usepackage{latexsym}
\usepackage{amssymb}
\usepackage{bm}
\usepackage[dvips]{graphics}
\usepackage{graphicx}
\usepackage{psfrag}
\usepackage{amsfonts,amsmath,amssymb,hyperref}
\usepackage{algorithm}
\usepackage{algorithmic}
\usepackage{dsfont,color,subfigure,setspace, cite}

\DeclareMathOperator*{\argmin}{arg\,min}

\usepackage{xspace}
\usepackage{bbm}

\newcommand{\Ac}{\mathcal{A}}
\newcommand{\Bc}{\mathcal{B}}

\newcommand{\Ec}{\mathcal{E}}

\newcommand{\Nc}{\mathcal{N}}

\newcommand{\Sc}{\mathcal{S}}

\newcommand{\Xv}{{\bf X}}
\newcommand{\Yv}{{\bf Y}}

\newcommand{\xv}{{\bf x}}
\newcommand{\yv}{{\bf y}}
\newcommand{\zv}{{\bf z}}
\newcommand{\uv}{{\bf u}}
\newcommand{\vv}{{\bf v}}
\newcommand{\sv}{{\bf s}}

\newcommand{\Uh}{{\hat{U}}}
\newcommand{\Vh}{{\hat{V}}}

\newcommand{\xh}{{\hat{x}}}

\newcommand{\xt}{{\tilde{x}}}

\def\a{\alpha}

\def\d{\delta}
\def\e{\epsilon}

\DeclareMathOperator\E{E}
\let\P\relax
\DeclareMathOperator\P{P}

\newcommand{\Bern}{\mathrm{Bern}}

\newcommand\ie{i.e.,\xspace}
\def\textiid{i.i.d.\@\xspace}
\newcommand\iid{\ifmmode\text{ i.i.d. } \else \textiid \fi}

\newcommand{\ind}{\mathbbmss{1}}

\newcommand{\xhv}{{\bf \hat{x}}}
\newcommand{\x}{{\bf x}}
\newcommand{\my}{{\bf y}}

\newcommand{\ex}{{\rm e}}
\newcommand{\y}{\yv }
\newcommand{\xhb}{\mathbf{\hat{x}}}
\newcommand{\w}{\mathbf{w}}
\newcommand{\xtb}{\mathbf{\tilde{x}}}
\newcommand{\q}{\mathbf{q}}
\newcommand{\qhb}{\mathbf{\hat{q}}}

\newcommand{\pb}{\mathbf{p}}

\newtheorem{definition}{Definition}

\newtheorem{theorem}{Theorem}
\newtheorem{lemma}{Lemma}

\newtheorem{corollary}{Corollary}
\newtheorem{remark}{Remark}
\newtheorem{proposition}{Proposition}
\newtheorem{example}{Example}
\begin{document}

\title{\LARGE \bf Minimum Complexity Pursuit \\ for Universal Compressed Sensing\footnote{This paper was presented in part at   Allerton Conference on Communication, Control and Computing, 2011 and at IEEE International Symposium on Information Theory,  Cambridge, MA, 2012.}}

\author{Shirin Jalali, Arian Maleki, Richard G. Baraniuk
\thanks{S. Jalali is with the Center for Mathematics of Information, California Institute of Technology, Pasadena, CA,
        {\tt\small shirin@caltech.edu}}%
\thanks{A. Maleki and R. G. Baraniuk are with the Digital Signal Processing group, Rice University, Houston, TX,
        {\tt\small $\{$arian.maleki, richb$\}$@rice.edu}}%
}

\maketitle

\newcommand{\p}{\mathds{P}}
\newcommand{\mb}{\mathbf{m}}   
\newcommand{\bb}{\mathbf{b}}

\begin{abstract}
The nascent field of compressed sensing is founded on the fact that high-dimensional signals with ``simple structure'' can be recovered accurately from just a small number of randomized samples.
Several specific kinds of structures have been explored in the literature, from sparsity and group sparsity to low-rankness.
However, two fundamental questions have been left unanswered, namely:
What are the general abstract meanings of  ``structure'' and ``simplicity''?
And do there exist universal algorithms for recovering such simple structured objects from fewer samples than their ambient dimension?
In this paper, we address these two questions.
Using algorithmic information theory tools such as the Kolmogorov complexity, we provide a unified definition of structure and simplicity.
Leveraging this new definition, we develop and analyze an abstract algorithm for signal recovery motivated by Occam's Razor.
Minimum complexity pursuit (MCP) requires just $O(3 \kappa)$ randomized samples to recover a signal of complexity  $\kappa$ and ambient dimension $n$.
We also discuss the performance of MCP in the presence of measurement noise and with approximately simple signals.
\end{abstract}


\section{Introduction}\label{sec:intro}
Compressed sensing (CS) refers to a body of techniques that undersample high-dimensional signals,  and yet recover them accurately by exploiting their intrinsic ``structure'' or ``compressibility" \cite{Donoho1, CaRoTa06}. This leads to more efficient sensing systems that have proved to be valuable in many applications, including cameras \cite{DuDaTaLaTiKeBa08}, magnetic resonance imaging (MRI) \cite{LuDoSaPa08} and radar \cite{BaSt07,HeSt09, AnMaBa12}, to name a few. While the promise of CS has been to undersample ``structured"  signals, its premise is still limited to specific instances of ``structure" such as sparsity and low-rankness. These notions are important in their own right. However, the concept of ``structure" and ``compressibility" is of course much more general than these specific instances. Several interesting extensions of sparsity and low-rankness that have been proposed in the last several years are testimonies to this claim \cite{RichModelbasedCS, ChRePaWi10, VeMaBl02, ReFaPa10, ShCh11, HeBa12, HeBa11, DoKaMe06}. The goal of this paper is to develop  a general and fundamental notion of structure for recovering signals from an undersampled set of linear measurements. In particular, we aim to answer the following question: Can we recover a given ``structured'' signal $\mathbf{x} \in \mathbb{R}^n$ from an undersampled set of linear measurements? Note that, unlike the other work in CS, the structure of the signal has not been specified. Therefore, to answer  this question we introduce  a {\em universal} notion of structure that distinguishes between ``structured'' and ``unstructured'' signals without employing any specific signal model. 

Towards this end, we use {\em  Kolmogorov complexity}, which is a measure of complexity for finite-alphabet sequences introduced by Solomonoff \cite{Solomonoff}, Kolmogorov \cite{KolmogorovC} and Chaitin \cite{Chaitin:66}. We argue that Kolmogorov complexity, if employed directly for the real-valued signals, is a restricted notion of complexity and does not cover well-knownn structures such as sparsity. Hence, based on Kolmogorov complexity, we define the \textit{Kolmogorov information dimension} (KID) of a real-valued signal as the growth rate of the complexity of its quantized version as the quantization becomes finer. Note that, similar to Kolmogorov complexity, KID is defined for individual sequences and is free of any signal modeling assumptions. Therefore, it provides a {\em universal} notion of structure. We prove that if  the KID of a signal is much smaller than its ambient dimension, then it  can be recovered  from fewer measurements than its ambient dimension. Furthermore, we show that KID of many well-studied structured signals  is small compared to their ambient dimensions, while the KID of well-known unstructured signals are ``close'' to their ambient dimensions.  

To demonstrate that approximate recovery of such structured signals is possible, we propose  the \textit{minimum complexity pursuit} (MCP) recovery algorithm. Based on Occam's razor \cite{occam}, MCP  approximates the simplest object (in the Kolmogorov complexity sense) that satisfies the measurement constraints. Roughly speaking, we prove that MCP is able to recover a signal with ``complexity'' $\kappa$ using no more than $3 \kappa$ measurements. Finally, we establish the robustness of MCP to noise on both the measurements and the signal.

The structure of the paper is as follows.  Section \ref{sec:def} describes  the notation used in the paper and introduces the KID. Section \ref{sec:contrib} summarizes our main contributions and their implications. Section \ref{sec:examp} bounds the KIDs of several popular classes of signals in CS. Section \ref{sec:related} makes a comparison of our work with the related papers in the literature.  Section \ref{sec:proofs} provides the proofs of our main results. Finally, Section \ref{sec:conclusion} concludes the paper.


\section{Background}\label{sec:def}
\subsection{Notation}
Calligraphic letters such as $\Ac$ and $\Bc$ denote sets. For a set $\Ac$, $|\Ac|$ and $\Ac^c$ denote its size and its complement, respectively. 
For a sample space $\Omega$ and  an event set $\Ac\subseteq \Omega$, $\ind_{\Ac}$ denotes the indicator function of the event $\Ac$. Boldfaced letters denote vectors. Throughout the paper, $\triangleq$ denotes equality by definition. For a vector $\xv \in \mathds{R}^n$, $x_i$, $\|\x\|_p\triangleq (\sum_{i=1}^n|x_i|^p)^{1/p}$, and $\|\x\|_{\infty}\triangleq\max_{i}|x_i|$ denote the $i^{\rm th}$ component, $\ell_p$ norm and $\ell_{\infty}$ norm of $\xv$, respectively. For $1\leq i\leq j \leq n$, $x_i^j \triangleq (x_i,x_{i+1},\ldots,x_j)$. Also, to simplify the notation, $x^j$  denotes  $x_1^j$. Uppercase letters are  used for both matrices and random variables, and hence their usage will be clear from the context. For  integer $n$, $I_n$ denotes the $n\times n$ identity matrix.

Let $\{0,1\}^*$ denote the set of all finite-length binary sequences, \ie $\{0,1\}^*\triangleq\cup_{n\geq 1}\{0,1\}^n$. Similarly, $\{0,1\}^{\infty}$ denotes the set of infinite-length binary sequences.

For a real number $x\in[0,1]$, let $[x]_m$ denote its $m$-bit approximation that results  from taking the first $m$ bits in the binary expansion of $x$. In other words, if   $x=\sum_{i=1}^{\infty}2^{-i}(x)_i$, where $(x)_i\in\{0,1\}$,  then
\[
[x]_m\triangleq\sum_{i=1}^{m}2^{-i}(x)_i.
\]
Similarly, for a vector $\x \in[0,1]^n$, define
\[
[\xv]_m\triangleq ([x_1]_m,\ldots,[x_n]_m).
\]
Throughout the paper, the basis of the logarithms is assumed to be $\ex$ unless otherwise specified.


\subsection{Kolmogorov complexity}\label{sec:kolm}
The prefix  Kolmogorov complexity of a finite-length binary sequence $\xv$ with respect to a universal computer ${\tt U}$ is defined as the minimum length over all programs that print $\xv$ and halt.\footnote{In Appendix \ref{app:kol}, we review some basic definitions of prefix Kolmogorov complexity. See \cite{book_vitanyi} for more details on the subject and also on the difference between prefix Kolmogorov complexity and its non-prefix version. }
For  $\x\in\{0,1\}^*$, let $K_{\tt U}(\xv)$ denote the Kolmogorov complexity of sequence $\xv$ with respect to the universal computer ${\tt U}$. Given an optimal universal computer ${\tt U}$ and any computer ${\tt A}$, there exists a constant $c_{\tt A}$ such that $K_{\tt U}(\x)\leq K_{\tt A}(\x)+c_{\tt A}$, for all strings $\x\in\{0,1\}^*$ \cite{book_vitanyi, cover}. This result is known as the {\em invariance theorem} in the field of algorithmic complexity. Note that the constant $c_{{\tt A}}$ is independent of the length of the sequence, $n$, and hence  can be neglected for sufficiently long $\xv$. As suggested in \cite{cover}, we drop the subscript ${\tt U}$, and let $K(\x)$ denote the Kolmogorov complexity of the binary string $\xv$ . \noindent For two finite alphabet sequences $\xv  $ and $\yv $, $K(\xv   \   |\ \yv )$ is defined as the length of the shortest program that prints $\xv  $ and halts, given that the universal computer ${\tt U}$ has access to the sequence $\yv $.\footnote{Note that $K(\xv \ | \yv)$ is often defined as $K(\xv   \ | \yv, \pb_\y)$ where $\pb_\y$ is the shortest program that generates $\yv $. This formulation provides symmetry in the definition of algorithmic mutual information. But we will not use this definition in this paper.}
 Similarly, the Kolmogorov complexity of an integer $n\in\mathds{N}$, $K(n)$, is defined as the Kolmogorov complexity of its binary representation. 
The following theorem summarizes some of the properties of the Kolmogorov complexity that will be used throughout the paper. Define 
\[
\log^*n \triangleq \lceil \log_2 n\rceil + 2\log_2 \max(\lceil \log_2 n\rceil ,1).
\]
\begin{theorem}[Properties of Kolmogorov complexity from \cite{book_vitanyi, cover}]\label{thm:properties}
Let  $\xv  , \yv $ be  binary strings of lengths $\ell(\x)$ and $\ell(\my)$, respectively. Furthermore, let  $m,n \in \mathds{N}$. The Kolmogorov complexity satisfies the following properties:
\begin{itemize}
\item[i.] $K(\x \, | \, \ell(\x) ) \leq \ell(\x) + c $,
\item[ii.] $K(\x,\my) \leq K(\x)+ K(\my)+c$,
\item[iii.] $K(\x \ | \; \my) \leq K(\xv)+c$,
\item[iv.] $K(\x) \leq K(\x \ | \ \ell(\x))+K(\ell(\x)) + c$,
\item[v.] $K(n) \leq  \log^* n+c$,
\item[vi.] $K(n+m) \leq K(n)+ K(m)+c$,
\end{itemize}
where $c$ is a constant independent of $\x, \my, n$ and $m$, but might be different from one appearance to another.
\end{theorem}
 While the proofs of different parts of this theorem can be found in \cite{book_vitanyi, cover}, for the sake of completeness, we present a brief summary of the proofs  in Appendix \ref{app:proof_thmprop}.
 
Kolmogorov complexity provides a  universal measure for compressibility of sequences.  It can be proved that an infinite length binary sequence $\xv$  is ``random'' if and only if there exists a constant $c$ such that
 \[
K(x_1, x_2, \ldots,x_n)> n-c
 \]
 for all $n$. (See  \cite{book_vitanyi} and its Theorem 3.6.1 for the exact definition of randomness and the proof of this result.)
 Furthermore, if the Kolmogorov complexity of $\xv$  is smaller than the ambient dimension, then it means that we can compress  $\xv$ (represent it with fewer bits); the encoder returns the shortest program that has generated $\xv$  and the decoder is the universal Turing machine that generates $\xv$,  from this short program.  
 
 \section{Problem statement}

\subsection{Compressed sensing versus compression}

Algorithmic information theory is mainly concerned with finding the shortest description of binary (or finite alphabet) sequences with respect to a universal computer. Similarly, in data compression the goal is to provide ``efficient'' representations of sequences, such that a decoder can recover them from their descriptions. However, in this paper we are interested in the problem of CS, where the goal is to reconstruct a signal $\x_o \in \mathds{R}^n$ from its lower dimensional linear projections $\yv_o = A \xv_o$, where $A \in \mathds{R}^{d \times n}$ with $d < n$. This problem has  two distinguishing features. First, since the system of equations is underdetermined, perfect reconstruction is not always possible. Therefore some knowledge of the structure of $\xv_o$ is required for recovering it from the measurements $\yv_o$. Second, the problem is different from the traditional problem of algorithmic information theory that considers the compression in terms of bits. Hence, this problem requires a new perspective on the Kolmogorov complexity of real-valued signals.

\subsection{Kolmogorov information dimension}

 Following the ideas in algorithmic information theory, one can consider the ``structure" of  a binary sequence to be the shortest program that generates it \cite{DoKaMe06}. The shorter the program, the more structured the signal.  Consider $\x_o \in [0,1]^n$, and define the Kolmogorov complexity of $\x_o$ as Kolmogorov complexity of the the binary sequence derived from  the concatenation of binary expansions of the components of $\x_o$. Using this definition, except for a set of measure zero, all  signals in $ [0,1]^n$ have infinite Kolmogorov complexity. Therefore, this notion does not capture many well-known structures for real-valused  signals such as sparsity. The first step to remedy this issue is to calculate the Kolmogorov complexity of a ``quantized'' version of $\xv_o $. For $\xv=(x_1,x_2,\ldots,x_n)\in [0,1]^n$, define the Kolmogorov complexity of $\xv$  at  resolution $m$ as 
\begin{eqnarray}\label{eq:quantizedKol}
K^{[\cdot]_m}(\xv)  \triangleq \inf_{\uv\in[0,1]^n} \left\{ K(\uv \ | \ n,\, m) \ | \  \|\xv- \uv \|_{\infty} \leq 2^{-m}\right\}.
\end{eqnarray}
We can provide an upper bound for $K^{[\cdot]_m}(\xv)$ by considering certain instances of $\mathbf{u}$. For example, $ \|\xv- [\xv]_m \|_{\infty} \leq 2^{-m}$, therefore,
\[
K^{[\cdot]_m}(\xv) \leq K([\xv]_m \ | \ m,n).
\]

Note that $K^{[\cdot]_m}(\xv)$ is defined as the Kolmogorov complexity of  the ``quantized'' version of $\xv$ conditioned on $m$ and $n$, because it is natural to assume that the encoder and decoder  have access to both the ambient dimension $n$ and the quantization level $m$. For most real valued signals this quantity goes to infinity as $m$ approaches infinity. But, the growth rate is proportional to $m$. Therefore, in this paper we consider a normalized version of the Kolmogorov complexity.

\begin{definition}
The {\em Kolmogorov information dimension (KID)} of  $(x_1, x_2, \ldots, x_n)\in[0,1]^n$ at resolution $m$ is defined as
\[
\kappa_{m,n}(\x) \triangleq \frac{K^{[\cdot]_{m}}(x_1,x_2, \ldots, x_n)}{m}.
\]
\end{definition} 
\noindent In general  the number of quantization levels $m$ may depend on the ambient dimension $n$. The division of $K^{[\cdot]_m}(\xv)$ by the resolution level $m$ ensures that for a fixed value of $n$ this quantity is always finite.

\begin{lemma}\label{lem:upperkol}
Let $\x \in [0,1]^n$. Then we have
\[
\kappa_{m,n}(\x) \leq n + \frac{c}{m},
\]
where $c$ is a positive constant independent of  $m$, $n$, and $\xv$ . In particular, 
\[
\lim \! \! \sup_{m \rightarrow \infty} \kappa_{m,n}(\x) \leq n.
\]
\end{lemma}

{\em Proof:}
We first note that
\begin{eqnarray*}
K^{[\cdot]_m}(\xv) &=& \inf_{\uv\in[0,1]^n} \left\{ K(\uv \; | \; n,\, m) \ | \  \|\xv- \uv \|_{\infty} \leq 2^{-m} \right\} \nonumber \\
 & \leq & K\left([\xv]_m|m,n\right).
\end{eqnarray*}
  Now, we derive an upper bound on $K([\xv]_m|n,m)$ by providing a program that describes $[\xv]_m$ conditioned on knowing $m$ and $n$. Consider the program that first explains the structure of the sequence as consisting of $n$ $m$-bit subsequences and then identifies the bits. Since the computer has access to $m$ and $n$, a constant number of bits (independent of  $m$ or $n$) is sufficient for specifying  the structure, and it then requires $mn$ more bits to specify each component $[x_i]_m$. Therefore,  overall 
\begin{align*}
\kappa_{m,n}(\x) & \leq \frac{K([x_1]_m, [x_2]_m, \ldots, [x_n]_m \, | \, m,n)}{m} \leq \frac{nm+ c}{m}. 
\end{align*}
The second part of theorem is a straightforward result of the first part. $\hfill \Box$

\begin{remark}\label{remark:1}
 Note that the existence of a finite upper bound on $K^{[\cdot]_m}(\x)$ ensures that the infimum in \eqref{eq:quantizedKol} is achieved. This is due to the fact that the number of sequences $(u_1, u_2, \ldots, u_n)$ that have $K(u_1,u_2, \ldots, u_n) \leq mn+c$ is finite. In the rest of the paper we denote the minimizing vector by $\phi_m(\x)$, i.e.,
 \begin{align}
 \phi_m(\x) \triangleq \arg \min_{\uv \in [0,1]^n}  \left\{ K(\uv \ | \ n,\, m) \ | \  \|\xv- \uv \|_{\infty} \leq 2^{-m}\right\}.\label{eq:def-of-phi-m}
 \end{align}
\end{remark}

The following examples clarify some of the  properties of the KID.

\begin{example}\label{ex:sparse}
(Sparse signals) Consider a $k$-sparse signal $\x \in [0,1]^n$.  That is, $\xv$ has at most $k$ nonzero coefficients. For any given $\d >0$, the KID of $\xv$ at resolution $m$, for large enough values of $m$, is upper bounded by $2k(1 + \delta)$. See Section \ref{sec:sparsity} for the proof of this claim.
\end{example}

\begin{example}\label{ex:lowrank}
(Low-rank matrices) Let $X$ denote a $M\times N$ real-valued matrix such that $\sigma_{\rm max}(X)\leq 1$.\footnote{As long as all the singular values are upper bounded by a constant the statement of this example holds. For the notational simplicity we choose $1$ as the upper bound for the singular values.} For any given $\delta>0$, the KID of $X$ at resolution $m$ is upper bounded by $r(M+N+1)(1+ \delta)$, for sufficiently large values of $m$. See Section \ref{sec:lowrank} for the proof of this claim.
\end{example}



%
%
Let $U[a,b]$ denote the uniform distribution between $a$ and $b$. Also, let $X \sim {\rm Bern}(p)$ represent a Bernoulli random variable with $\P(X=1) = 1- \P(X=0)=p$. The following proposition lets us construct the third example that represents an unstructured signal. 

\begin{proposition}\label{prop:uniform}
Let $\{X_i\}_{i=1}^{\infty} \overset{i.i.d.}{\sim} U[0,1]$. Then, for any $n\geq 1$, 
\[
\lim_{m\to\infty}\frac{1}{mn}K^{[\cdot]_m}(X_1, X_2, \ldots X_n) = 1
\]
in probability.
\end{proposition}
{\em Proof:}
For $i\in\{1,2,\ldots\}$, let $X_i=\sum_{j=1}^{\infty} (X_i)_j2^{-j}$, where $(X_i)_j\in\{0,1\}$. Then $\{(X_i)_j\}_{j=1}^{\infty} \overset{i.i.d.}{\sim} \Bern(1/2)$ \cite{Marsaglia12}. Let $U^n\triangleq\phi_m(X^n)$. Since $|U_{i}-X_i|\leq2^{-m}$, then,   for $j<m-1$, $(X_i)_j = (U_i)_j$. Therefore,
\begin{align}\label{eq:Kl1}
\frac{K(U^n\ | \ m, \, n )}{m}  & \geq  \frac{K({\{((U_i)_1,\ldots,(U_i)_m) \}_{i=1}^{n}} \ | \ m, \, n )-c}{m} \nonumber \\
& = \frac{K({\{((X_i)_1,\ldots,(X_i)_m) \}_{i=1}^{n}} \ | \ m, \, n )-c}{m}.
\end{align}
Theorem 14.5.3 in \cite{cover} states that  the normalized Kolmogorov's complexity of a sequence of i.i.d. $\Bern(1/2)$ bits converges to $1$ in probability. In other words,
\begin{align}
\lim_{m\to\infty}{K(\{(X_i)_1,(X_i)_2,\ldots,(X_i)_m\}_{i=1}^n\, | \, m, n) \over mn}= 1,\label{eq:K1}
\end{align}  
in probability.  Therefore, combining \eqref{eq:Kl1}, Lemma \ref{lem:upperkol} and \eqref{eq:K1} yields the desired result. $\hfill \Box$ \\

\begin{example}\label{example:uniform}
If the random variables $\{X_i\}_{i=1}^{n} \overset{i.i.d.}{\sim} U[0,1]$, then
\[
\lim_{m \rightarrow \infty} \frac{K^{[\cdot]_m}(X_1, X_2, \ldots, X_n)}{m} = n
\]
in probability. The proof follows directly from Proposition \ref{prop:uniform}.
\end{example}

\noindent These examples demonstrate that, at least in cases where the ambient dimension is fixed and the quantization levels grow without bound, the KID is much smaller than the ambient dimension for the two well-known structured signals in Examples \ref{ex:sparse} and \ref{ex:lowrank}, and is equal to the ambient dimension for the unstructured signal in Example \ref{example:uniform}. We present more examples of structured signals and the corresponding upper bounds on their KID in Section \ref{sec:examp}. 

%
%

\vspace{.1cm}

\subsection{Minimum complexity pursuit}
Consider the problem of recovering a structured real-valued signal $\xv_o=(x_{o,1},x_{o,2},\ldots)$ with   $\kappa_{m,n}(x_o^n) = O(n^{1-\alpha})$, for some $\alpha>0$ and proper choice of $m$, from an underdetermined set of linear equations $\yv_o = A\xv_o $, where $\yv_o \in \mathds{R}^d$ and $d<n$. We follow Occam's Razor and among all the solutions of $\yv_o = A\xv_o $, seek the solution that has the minimum complexity, i.e., 
\begin{eqnarray}
&&\arg \min \quad K^{[\cdot]_m}(\xv)\nonumber \\
&&{\rm s.t.}\quad \ \  \;\;\;\; A\xv = \yv_o.\label{eq:alg}
\end{eqnarray}
We call this algorithm {\em minimum complexity pursuit} or MCP. MCP has a free parameter $m$ whose effect on the performance of the algorithm will be discussed in detail later.  We will show that MCP can recover $\xv_o $ from fewer measurements than the ambient dimension of the signal. This result extends the scope of CS from the class of sparse signals or the class of low-rank matrices to the class of all signal with small KID.

In this paper we ignore the practical issues of approximating the MCP algorithm. In an independent work, \cite{BaDu11, BaDu12} have considered a practical version of this algorithm and provided promising results in that direction. Note that the model that is considered in \cite{BaDu11, BaDu12} is restricted to the stochastic signals that are drawn from an unknown distribution. Such restrictions might be required for obtaining practical algorithms. Further investigation of the practical issues is left  as an avenue for future research.


\section{Our contributions}\label{sec:contrib}

\subsection{Recovery in the noiseless setting}\label{sec:A}

Suppose that  $A\in\mathds{R}^{d\times n}$, $\xv_o\in\mathds{R}^n$ and   $\yv_o=A\xv_{o}$. 
We are interested in recovering $\xv_o$ from its linear measurements $\yv_o$. Let $\xhb_o=\xhb_o(\y_o,A)$ denote the output of \eqref{eq:alg} to the inputs $\yv_o$ and $A$.  The following theorem states that having enough number of measurements, \eqref{eq:alg} succeeds in recovering $\xv_o$. 

\begin{theorem}\label{thm:1}
Let $\x_o\in[0,1]^n$, and let $\kappa_{m,n}=\kappa_{m,n}(\x_o)$ denote the information dimension of $\x_o$ at resolution $m$.  Also, let $\xhv_{o}$ denote the solution of  \eqref{eq:alg} to $\yv_o=A\x_o$, where $A_{ij}$ are i.i.d. $\Nc(0,1)$. Then,   for any $t\in(0,1)$, we have
\begin{align*}
\P &\left(\| \x_{o}-\xhb_{o}\|_2>  \left({1\over\sqrt{1-t}}\left(\sqrt{n\over d}+2\right) +1\right){\sqrt{n}\over 2^m}\right) \nonumber \\
&\leq 2^{ \kappa_{m,n} m} {\rm e}^{\frac{d}{2} (t +\log(1-t) )} + {\rm e}^{- \frac{d}{2} }. 
\end{align*}
\end{theorem}

The proof is presented in Section \ref{sec:proofthm1}. Note that $\kappa_{m,n}$ in Theorem \ref{thm:1} is both a function of $m$ and $n$. Next, we consider several  interesting corollaries of this theorem for high dimensional problems.

\begin{corollary}\label{cor:noiseless_unnormerror}
Assume that $\x_o\in[0,1]^{n}$ and $m = \lceil \log n \rceil$. Let $\kappa_n\triangleq \kappa_{m,n}(x_o^n)$ and  $d = \lceil \kappa_n \log n\rceil$. Assume that $d\leq n$. Then, 
\[
\P\Big(\|\x_{o}-\xhb_{o}\|_2> {20\over \sqrt{d}} \Big) < 2\ex^{-{d\over 2}}.
\]
\end{corollary}
{\em Proof:}
For $m= \lceil \log n \rceil$, $2^{-m}\sqrt{n}\leq n^{-0.5}$. Choosing $t=0.965$, we get
\begin{align*}
\left((1-t)^{-0.5}\left(\sqrt{nd^{-1}}+2\right) +1\right)2^{-m}\sqrt{n} &\leq {1\over \sqrt{(1-t)d}}+{1+2(1-t)^{-0.5}\over \sqrt{n}} \nonumber\\
&\leq {20\over \sqrt{d}},
\end{align*}
where the last step follows since $d\leq n$.
Therefore, by Theorem \ref{thm:1},
\begin{align*}
\P\left(\|x_{o}^n-\xh_{o}^n\|_2^2>  {20 \over \sqrt{d}} \right)  &\leq 2^{ \kappa_{m,n} m} {\rm e}^{\frac{d}{2} (t +\log(1-t) )} + \ex^{-{d\over 2}}\nonumber\\
&\leq 2 {\rm e}^{- \frac{d}{2}}.
\end{align*}
$\hfill \Box$

According to Corollary \ref{cor:noiseless_unnormerror}, if the complexity of the signal is less than $\kappa$, then the number of linear measurements required for its asymptotically perfect recovery is roughly speaking on the order of $\kappa \log n$. In other words, the number of measurements is proportional to the complexity of the signal and only logarithmically proportional to its ambient dimension.

\begin{corollary}\label{cor:noiseless_normerror}
Assume that $\x_o\in[0,1]^{n}$, $m =2 \lceil \log n \rceil$ and $\kappa_n=\kappa_{m,n}(x_o^n)$. Then, for $d = 3 \kappa_n $, we have
\[
\P\left(\|\x_{o}-\xhv_{o}\|_2> {4\over d} \right) <\ex^{-0.1 \kappa_n\log n}+\ex^{-0.5d} .
\]
\end{corollary}

{\em Proof:} Setting $t=1-{1\over n}$, $m = 2\lceil \log n \rceil$, and $d=\lceil 3 \kappa_n \rceil$, we have 
\begin{align*}
2^{ \kappa_{m,n} m} {\rm e}^{\frac{d}{2} (t +\log(1-t) )}&\leq 2^{2 \kappa_{n} \log n} {\rm e}^{1.5 \kappa_n(1-\log n )}\nonumber\\
&<\ex^{-0.1 \kappa_n\log n},
\end{align*}
for $n$ large enough. Also,
\begin{align}
 \left((1-t)^{-0.5}\left(\sqrt{nd^{-1}}+2\right) +1\right)2^{-m}\sqrt{n} &\leq {1+\sqrt{n}(2+\sqrt{n/d})\over n\sqrt{n}}\nonumber\\
 &<{3\over n} + {1\over \sqrt{nd}}
 <{4\over d}.
\end{align}
$\hfill \Box$


It is worth noting that, while $m$ is set to $O(\log n)$ in Corollaries  \ref{cor:noiseless_unnormerror} and \ref{cor:noiseless_normerror}, it can be considered as a free parameter of the MCP algorithm. Theorem \ref{thm:1} describes the trade-off of the parameters. If we fix all the other parameters in Theorem \ref{thm:1}, then increasing $m$ is equivalent to decreasing the reconstruction mean square error. But also it decreases the probability of correct recovery. 

\subsection{Recovery in the presence of Gaussian noise in measurements}\label{ssec:stochnoise}
In the previous section, we considered the case of recovering low-complexity  signals from their noise-free linear measurements. In this section, we extend these results to the case of noisy measurements, where $\yv_o = A\xv_o  + \w$, with $\w \sim \Nc(0, \sigma^2 I_d)$. Assuming that the complexity of the signal is known at the reconstruction stage, we consider the following reconstruction algorithm:
\begin{eqnarray}\label{eq:recover_noisy}
&& \arg \min \;\;\; \|A\x-\yv_o\|_2, \nonumber \\
&&{\rm s.t.}\;\;\;\;\;\;\;\;\; \ K^{[\cdot]_{m}}(\x) \leq \kappa_{m,n} m.
\end{eqnarray}
 Note that $\kappa_{m,n} m$ is an upper bound on the Kolmogorov complexity of $\xv_o $ at resolution $m$. We call this algorithm {\em low-complexity least squares} (LLS). Our quest in this section is to find the number of measurements required to make the LLS algorithm specified by \eqref{eq:recover_noisy} robust to noise. 

\begin{theorem} \label{thm:noisysetting}
Consider $\x_o\in[0,1]^{n}$. Let  $m=\lceil\log n\rceil $, $\kappa_n=\kappa_{m ,n}(x_o^n)$ and $d=\lceil8r\kappa_{n}m\rceil$,  where $r>1$. Also let  $\xhv_o$ denote the solution of LLS to input $\yv_o=A\xv_o+\w$, where $\{A_{ij}\}_{i,j}$ are i.i.d. distributed as $\Nc(0,1)$ and $\{w_i\}_i$ are i.i.d.~distributed as $\Nc(0,\sigma^2)$. Then,
\begin{align}\label{eq:mse_noisy}
\P\left(\|\x_{o}-\xhv_{o}\|_2^2 >  {9\sigma^2\over r} \right) <  6\ex^{-0.01d}+ \ex^{-0.3m\kappa_n},
\end{align}
for $d$ and $n$ large enough and $\sigma>0$.
\end{theorem}

The proof is presented in Section \ref{sec:proofthmnoisy}. 
\begin{remark} Note that, since the elements of the matrix $A$ are i.i.d. $\Nc(0,1)$, as the ambient dimension $n$ grows, so does the signal-to-noise (SNR) ratio per  measurement. In order to have fixed SNR ratio per measurement, one can draw  the elements of $A$ i.i.d.\ from $\Nc(0, 1/n)$. In this case, it is not difficult to see that the normalized mean square error $\|\x_{o}-\xhv_{o}\|_2^2/n \leq {9\sigma^2 \over r}$,  in probability. \end{remark}



\subsection{Recovery in the presence of deterministic noise} \label{sec:result_deterministic}
Consider again the measurement system we introduced in the last section: $\yv_o = A\xv_o + \w $, where $\w$ represents measurement noise. Unlike the previous section, assume that the noise is deterministic and has bounded $\ell_2$-norm, i.e., $\|\w\|_2 \leq e$. This type of noise provides a good model for quantization noise on the measurements, among other practical nonidealities. Note that unlike the case of stochastic noise, deterministic noise can be adversarial. We prove that the LLS algorithm  \eqref{eq:recover_noisy} provides a sufficiently accurate estimate of $\xv_o $ even in the presence of such noise.

\begin{theorem}\label{thm:3}
Let $\x_o=(x_{o,1},\ldots,x_{o,n})\in[0,1]^n$ and $\yv_o = A\xv_o + \w$, where $\|\w\|_2 \leq e$. Let $\kappa_{m,n}=\kappa_{m,n}(\x_o)$ denote the information dimension of $\x_o$ at resolution $m$.  Then, for any $t\in(0,1)$, we have 
\begin{eqnarray}
\lefteqn{\P \left(\|\x_{o}-{\bf \xh}_{o}\|_2> \left({1\over \sqrt{1-t}}\Big(\sqrt{n\over d}+2\Big) +1\right)2^{-m}\sqrt{n}+ \frac{e}{\sqrt{(1-t)d}}\right)} \nonumber \\
&\leq& 2^{ \kappa_{m,n} m} {\rm e}^{\frac{d}{2} (t +\log(1-t) )} + {\rm e}^{- \frac{d}{2} }. \hspace{4cm} \nonumber
\end{eqnarray}
\end{theorem}
Since the proof of this theorem is very similar to the proof of Theorem \ref{thm:1}, it is not included in the paper. Here the probability of accurate recovery is the same as in Theorem \ref{thm:1}, and under similar conditions this probability converges to one. The reconstruction error has two terms. The first term is again similar to Theorem \ref{thm:1} and under similar conditions converges to zero. The second term in the error, $\frac{e}{\sqrt{(1-t) d}}$,  is due to the noise in the measurements. As the number of measurements increases, $\frac{e}{ \sqrt{(1-t) d}}$ converges to zero. This is due to the fact that  since $A_{i,j} \sim \Nc(0,1)$ as we increase the number of measurements, the energy of the signal per measurement is fixed. But since the total amount of energy of the noise is considered to be constant the average noise per measurement decreases by $1/\sqrt{d}$. 
 
\subsection{Recovery of approximately low-complexity signals} \label{sec:result_approx}

In Sections  \ref{sec:A}-\ref{sec:result_deterministic},  we  considered recovering  ``low-complexity'' signals from their linear (noisy or noise-free) projections. However, most applications feature signals that are not of exactly low-complexity but rather are ``close'' to low-complexity signals. An example is the class of power-law ``compressible'' signals,  discussed in Section \ref{sec:power}, which are a popular model in the CS literature and are more realistic than sparse signal models.
 In this section, we discuss this more general setting. Assume that the original signal $\xv_o $ is  not low-complexity but is close to the low-complexity signal $\xtb$, i.e., $\|\xv_o - \xtb\|_2 \leq \epsilon_n$ with $\epsilon_n = o(1)$. Again, let $\yv_o=A\xv_o$. Consider the following reconstruction algorithm for recovering $\xv_o $ from its noisy linear measurements $\yv_o$:
\begin{eqnarray}
&&\min\quad  \|\yv_o-A\x\|_2^2 \nonumber\\
&&{\rm s.t.}\quad \ \ K^{[\cdot]_m}(\xv) \leq \kappa_{m,n} m .\label{eq:alg_model_mismatch}
\end{eqnarray}

\noindent Assume that $A\in\mathds{R}^{d\times n}$ and $A_{ij}$ are i.i.d.\ $\Nc(0,1)$. Let $\xhb_o=\xhb_o(\yv_o,A)$ denote the solution of \eqref{eq:alg_model_mismatch}.

\begin{theorem}\label{thm:4}
Assume that there exists $\xtb_o\in\mathds{R}^n$ such that $\|\xv_o -\xtb_o\|_2 \leq \epsilon_n$, and ${K^{[\cdot]_m}(\xtb_o)}\leq \kappa_{m,n}m$. Let $\yv=A\xv_o$, where $A$ is a $d\times n$ matrix with i.i.d.\ $\Nc(0,1)$ entries, and let $\xhb_o$ denote the minimizer of \eqref{eq:alg_model_mismatch}. Then, for any $0<t<1$, 
\begin{align}
\P& \Big(\|\x_{o}-\xhb_{o}\|_2> {1\over \sqrt{1-t}}(\sqrt{n\over d}+2)(2^{-m}\sqrt{n}+2\e_n) +2^{-m}\sqrt{n}\Big) \nonumber \\
&\leq 2^{ \kappa_{m,n} m} {\rm e}^{\frac{d}{2} (t +\log(1-t) )} + {\rm e}^{- \frac{d}{2} }. \hspace{4cm} \label{eq:8}
\end{align}

\end{theorem}
The proof is presented in Section \ref{sec:proof2}. There are two main error terms in \eqref{eq:8}. The first one  is the reconstruction error due to the quantization performed in the calculation of Kolmogorov complexity. The second term is due to the fact that the signal $\xv_o $ is not of exactly low-complexity. The following corollary simplifies the statement of the theorem for some special useful cases.

\begin{corollary}\label{cor:approxsparse}
Consider $\x_o \in [0,1]^n$ and assume that there exists  $\xtb_o \in [0,1]^{n}$, such that  $\|\x_o-\xtb_o\|\leq\e_n$. Let $m= \lceil \log n\rceil$ and $\kappa_n=\kappa_{m_n,n}(\xtb_o)$, $d = \lceil \kappa_n \log n \rceil$, $\yv_o=A\x_o$, where $A$ is a $d\times n$ matrix with i.i.d.\ $\Nc(0,1)$ entries, and $\xhv_o=\argmin_{K^{[\cdot]_m}(\xv) \leq \kappa_{n} m}\|\yv_o-A\xv\|$. Then,
\[
P\Big(\|\x_o - \xhv_o\|_2^2 > 25\e_n\sqrt{n\over d}\;\Big) < 2\ex^{-0.5d},
\]
for $d<n$ large enough.
\end{corollary}
{\em Proof:}
Setting $t=0.965$ 
\begin{align*}
 2^{ \kappa_{m,n} m} {\rm e}^{\frac{d}{2} (t +\log(1-t) )} + {\rm e}^{- \frac{d}{2} } &< 2^{d} {\rm e}^{\frac{d}{2} (0.965 +\log 0.035 )} + {\rm e}^{- \frac{d}{2} } \nonumber\\
 &< 2\ex^{-0.5d}. 
\end{align*}
Also, for the same value of $t$ and $m_n=\lceil\log n\rceil$,
\begin{align*}
{1\over \sqrt{1-t}}(\sqrt{n\over d}+2)(2^{-m}\sqrt{n}+2\e_n) +2^{-m}\sqrt{n}& < 6(\sqrt{n\over d}+2)({1\over\sqrt{n}}+2\e_n) +{1\over \sqrt{n}}\nonumber\\
&< 25\e_n\sqrt{n\over d},
\end{align*}
where at the last step we have assumed that $d<n$, and both of them are large enough.
$\hfill \Box$

\subsection{Other measurement matrices}
For the sake of clarity, the results presented so far have focused on i.i.d. Gaussian measurement matrices. However, the results can be extended to the more general class of i.i.d.\ subgaussian matrices.

\begin{definition}
A random variable $X$ is called {\em subgaussian}  if and only if there exist two constants $c_1, c_2>0$ such that 
\[
\P(|X|> t) \leq c_1 {\rm e}^{-c_2 t^2}.
\]
Such a random variable is denoted by $ {\rm SG}(c_1,c_2)$.
\end{definition}

Our goal in this section is to show how our results can be extended to the problem of CS with i.i.d. subgaussian measurement matrices. Our main conclusion is that the results presented for Gaussian matrices continue to hold for subgaussian matrices except for slight changes in the constants. However, as will be discussed later in Section \ref{sec:proofsubgauss}, the proof techniques are different from those for Gaussian matrices. To show these differences we extend the result of Theorem \ref{thm:1} to subgaussian matrices. Similar arguments can be used for other extensions. As before we consider the problem of recovering $\xv_o $ from linear measurements $\yv_o = A \xv_o $, where the elements of the matrix are i.i.d.\ ${\rm SG}(c_1,c_2)$. 

\vspace{.1cm}

\begin{theorem}\label{thm:sub-Gaussian}
Let $\x_o\in[0,1]^{n}$. For integers $m$ and $n$, let $\kappa_{m,n}=\kappa_{m,n}(\x_o)$. Assume that $\yv_o=A\x_o$, where $A$ is a $d\times n$ matrix, such that its entries are i.i.d.~distributed as ${\rm SG}(c_1,c_2)$, and $\E[A_{ij}]=0$ and $\E[A_{ij}^2]=1$.    Then, there exist three constants $c'_1$, $c'_2$, and $c_3$ depending only on $c_1$ and $c_2$ such that for any $1- \frac{c_3}{c_2}<\tau <1$ 
\begin{eqnarray}
\lefteqn{\P \left(\|\x_{o}-\xhb_{o}\|_2>  ({ \tau^{-1}(\sqrt{ (c'_2+1){n}/{d}}+1) +1} )2^{-m}\sqrt{n}\right)} \nonumber \\
&\leq& 2^{2 \kappa_{m,n} m} {\rm e}^{- \frac{dc_2^2(\tau^2-1)^2}{16 c_3}} +{\rm e}^{- c'_1 n}. \hspace{4cm} \nonumber
\end{eqnarray}
\end{theorem}

Theorem \ref{thm:sub-Gaussian} shows that, by choosing $m = \lceil \log n \rceil$,    $O(\kappa_{m,n} \log n)$ measurements remain sufficient for asymptotically  accurate recovery. But, as expected, the constants might be different from those in Theorem \ref{thm:1}.

\subsection{Discussion}\label{sec:discussion}
The LLS algorithms  proposed in  \eqref{eq:recover_noisy} and \eqref{eq:alg_model_mismatch}, corresponding to the cases  when noise is present either in the signal or in the measurements, both assume the knowledge of  an upper bound on the complexity of the signal. While such knowledge might be available or estimated in some applications, in many cases it is not straightforward to acquire it. In those cases, one might change the formulation of the MCP as follows:
\begin{eqnarray}
&&\argmin\quad K^{[\cdot]_m}(\x)\nonumber \\
&&{\rm s.t.}\quad \ \  \;\;\;\; \|A\x - \y_o\|_2 \leq z_n.\label{eq:MCP-zn}
\end{eqnarray}
We call this new algorithm {\em relaxed MCP} or R-MCP. In this new optimization problem the challenge is to set parameter $z_n$ properly. The value of this parameter should be set according to the noise level present in the system. For instance, if we employ $z_{n} = (\sqrt{n}+ (t+1)\sqrt{d})\epsilon_n$ and $z_n = e$ for the approximately low-complexity signals case (corresponding to Section \ref{sec:result_approx}) and exactly sparse signal in the presence of deterministic noise (corresponding to Section \ref{sec:result_deterministic}), respectively, then we  obtain results that are exactly the same as those stated in Theorems \ref{thm:3} and \ref{thm:4}. Since the proofs are very similar to the proofs of Theorems \ref{thm:3} and \ref{thm:4}, we skip them here. In the case of stochastic noise (corresponding to Section \ref{ssec:stochnoise}), it is not clear if this new formulation  provides a bound similar to Theorem \ref{thm:noisysetting}. This problem is deferred to future research.


\section{Kolmogorov dimension of certain classes of functions}\label{sec:examp}

In this section, we explore the implications of our results for several signal classes to which CS has been successfully applied. We show that the number of measurements MCP requires for the accurate recovery is within the same order of the other well-known recovery algorithms. To achieve this goal we need to calculate KID for certain signals. 
 It is well known that the Kolmogorov complexity of a sequence is not computable (See \cite{cover}, Section 14.7). However, it is often possible to provide upper bounds on the Kolmogorov 
 complexity. In this section, we consider several standard classes of functions and provide upper bounds on
 their KID. Based on these upper bounds, one can use Theorems \ref{thm:1} and \ref{thm:4}
 to calculate the number of  linear measurements required by the MCP to recover them.  These examples demonstrate the connection between the results of Section \ref{sec:contrib} and the
 CS framework explained in the Introduction.

\subsection{Sparse signals} \label{sec:sparsity}
A class of signals that has played a key role in CS is the class of $k$-sparse signals.  The following proposition provides an upper bound on the KID of such signals.

\begin{proposition}\label{prop:sparse}
Let the signal $\xv_o =(x_{o,1}, x_{o,2}, \ldots,x_{o,n})$ be $k$-sparse, \ie $\|\xv_o\|_0\leq k$. Then
\[
\kappa_{m,n}(\xv_o ) \leq  k+ \frac{nh({k\over n})+0.5\log n+c}{m}.
\]
\end{proposition}
{\em Proof:}
 Consider the following program for describing $[\xv_o ]_m$. 
First, use a program of constant length to describe the structure of the signal as ``sparse''  and the ordering of the rest of information, and the length of the sequence and the resolution.\footnote{Note that in calculating the information dimension we assume that $n$ and $m$ are given to the universal computer. Otherwise we would need $\log^*n$ and $\log^*m$ bits to describe them to the machine.} Next, spend $nh({k\over n})+0.5\log n+c'$  bits, where for $\a\in[0,1]$, $h(\a)\triangleq -\a\log_2\a-(1-\a)\log_2(1-\a)$,  to code string of length $n$ that contains the locations of the $k$ non-zero elements \cite{cover}. Finally, use
$km$ more bits to describe the quantized magnitudes of the non-zero coefficients. Therefore, overall, we have
\begin{eqnarray*}\label{eq:kc_poly}
\lefteqn{K^{[\cdot]_{m}}(x_{o,1},x_{o,2},\ldots,x_{o,n}) }\nonumber \\
 &\leq& km+ {nh\Big({k\over n}\Big)+0.5\log n+c},
\end{eqnarray*}
where $c$ is a constant independent of $\xv_o$, $m$ and $n$. 
$\hfill \Box$

In most of our analysis in this paper we consider the case of $m = \lceil \log n\rceil$. It is straightforward to confirm that in this case, for $n,k$ sufficiently large and $k \ll n$ we have 
\[
\kappa_{m,n}(\xv_o ) \leq k+{n\over \log n}h\Big({k\over n}\Big)+1 \leq 2k(1+ \delta),
\] 
where $\delta$ is a small fixed number. It is straightforward to plug this upper bound in Corollary \ref{cor:noiseless_normerror} and prove that, for large values of $n$, $6k(1+ \delta)$ measurements are sufficient for the ``successful'' recovery of $k$-sparse signals. This is still larger than $2k$ measurements required by the $\ell_0$ minimization. However, the source of the discrepancy is not clear to the authors at this point.

\subsection{Power law compressible signals}\label{sec:power} While sparse signals have played an important role in the theory of CS, it is well-known 
that they rarely occur in practice. More accurate models assume that either the  signal's coefficients  decay at a specified rate, or the signal belongs to an $\ell_p$ ball with $p<1$  \cite{Donoho1}, i.e., the signal belongs to the set
\[
\mathcal{B}_p^n \triangleq \left\{ \x \in \mathds{R}^n  \ : \ \| \x\|_p  \leq 1 \right\}.
\]
For  $\x_o\in \mathcal{B}_p^n$, let $(x_{o,(1)},x_{o,(2)}, \ldots,x_{o,(n)})$ denote the permuted version of $\x_o$ such that $|x_{o,(1)}|\geq |x_{o,(2)}|\geq \ldots \geq |x_{o,(n)}|$. It is straightforward to show that $|x_{o,(i)}| \leq i^{-\frac{1}{p}}$, i.e., it is power law compressible. Therefore, if we just keep the
$k$ largest coefficients of this signal and set the rest to zero, the resulting $k$-sparse vector $\xtb_o$ satisfies: $\|\x_o- \xtb_o\|_2 \leq  k^{-\frac{1}{p}+\frac{1}{2}} $. In Section \ref{sec:sparsity}, we derived an upper bound for the KID of $\tilde{\x}_o$. Proposition \ref{prop:6} follows from this bound and Corollary \ref{cor:approxsparse}.

\begin{proposition}\label{prop:6}
Let $\x_o \in \Bc_p^{n}$,   $\y_o=A\x_o$, where $A$ is a $d\times n$ random matrix with i.i.d.\ $\Nc(0,1)$ entries. Set $d = \lceil 3n^{p/2}\log n \rceil$. Let  $\xhv_o$ denote the minimizer of \eqref{eq:alg_model_mismatch} with $m = \lceil \log n \rceil$ and $\kappa_{m,n} = 3n^{p/2}$. Then,
\[
  \P\Big(\|\x_{o}-\xhv_{o}\|_2> {7\over \sqrt{\log n}}\Big) \leq 2 {\rm e}^{- 0.5d },
 \] 
 for sufficiently large $n$.
\end{proposition}
{\em Proof:}
 Let $\xtb_o$ denote the $k$-sparse approximation of $\x_o$ derived by keeping the $k = n^{p/2}$ largest coefficients of $\x_o$, and setting the rest to zero. Then, $\|\x_o-\xtb_o\|_2\leq \e_n= n^{-\frac{1}{2}+\frac{p}{4}} $. According to Proposition \ref{prop:sparse}, for $n$ large enough, the KID of $\xtb_o$ at resolution $m = \log n$ is upper bounded by 
 \[
k+ \frac{nh({k\over n})+0.5\log n+c}{\log n} <2k(1 + \d),
 \] 
 where $\d>0$, can be made arbitrary small for $n$ large enough.
 By setting $\delta = 0.5$ we obtain $\kappa_{m,n}(\xtb_o) \leq 3 n^{\frac{p}{2}}$. Also, for  $t= 0.965$,
 \begin{align}
 {1\over \sqrt{1-t}}(\sqrt{n\over d}+2)(2^{-m}\sqrt{n}+2\e_n) +2^{-m}\sqrt{n}  <{7\over \sqrt{ \log n}},
\end{align}
for $d<n$ large enough. Therefore, Theorem \ref{thm:4}, yields the desired result.
$\hfill \Box$
 
 It is interesting to note that, as the power $p$ decreases, the number of measurements required for successful recovery decreases.

\subsection{Piecewise polynomial functions} \label{sec:polynomial}

 Let ${\rm Poly}_N^Q$ denote the class of  piecewise polynomial functions $f:[0,1] \rightarrow [0,1]$  with at most $Q$  singularities\footnote{A singularity is a point at which the  function is not infinitely differentiable.} and maximum degree of $N$. For $f\in{\rm Poly}_N^Q$, let  $(x_{o,1}, x_{o,2}, \ldots,x_{o,n})$ be the samples of $f$ at \[0, {1 \over n}, \ldots,{n-1\over n}.\]  Let $\{a_i^{\ell}\}_{i=0}^{N_{\ell}}$ denote the set of coefficients of the $\ell^{\rm th}$ polynomial of $f$, where $N_{\ell}\leq N $ denotes its degree. For the notational simplicity, we assume that the coefficients of each polynomial belong to the $[0,1]$ interval and that $\sum_{i=0}^{N_{\ell}} a^{\ell}_i <1$, for every $\ell$. Define 
\[
\mathcal{P} \triangleq \left\{\x_o \in \mathds{R}^n \ | \ x_{o,i} = f(i/n), \ f \in {\rm Poly}_N^Q \right\}.
\]

\begin{proposition}\label{prop:polynomial}
For every signal $\x_o \in \mathcal{P}$, we have
\begin{eqnarray*}
k_{m,n}(\x_o) &\leq&  (Q+1)(N+ 1)+\frac{(Q+1)(N+ 1)\lceil\log_2(N+1)\rceil}{m}\nonumber \\
                 &&+~\frac{\log^*n+ \log^*N + \log^* k + Q\log^*n+ c_1+c_2}{m}.
\end{eqnarray*}
\end{proposition}

{\em Proof:}
Consider the following program for describing the quantized version of$\xv_o$. The code  first specifies the signal model as samples of a ``piecewise polynomial'' function with parameters $(n,Q,N)$. This requires $\log^*N + \log^* Q + c$ bits. Then, for each singularity point, the code first specifies  the largest sampling point $i/n$ that is smaller than it. Since there are at most $Q$ singularity points, describing this information requires at most $Q \log^* n$ bits. The next step is to describe the coefficients of each polynomial. Using an $m'$-bit uniform quantizer for each coefficient, the induced error is bounded as
              \begin{align}
              \left|\sum_{i=0}^{N_{\ell}} a^{\ell}_i t^n-\sum_{i=0}^{N_{\ell}} [a^{\ell}_i ]_{m'} t^n \right| &\leq \sum_{i=0}^{N_{\ell}} |a^{\ell}_i- [a^{\ell}_i]_{m'}|\nonumber\\
              & \leq (N_{\ell}+1) 2^{-m'}\leq (N+1) 2^{-m'}. 
              \end{align}
To ensure reconstructing the samples at resolution $m$, we require  $(N+1) 2^{-m'}<2^{-m}$.  Therefore, to describe the  coefficients of the polynomials, at most, $(Q+1)(N+1) (m+\lceil\log_2(N+1)\rceil)$ extra bits are required. Hence, overall, it follows that
               \begin{align} \label{eq:kc_pp}
               {K^{[\cdot]_{m}}(x_{o,1},x_{o,2},\ldots,x_{o,n}) \over m}\leq&
                 (Q+1)(N+ 1)+\frac{(Q+1)(N+ 1)\lceil\log_2(N+1)\rceil}{m}\nonumber \\
                 &+\frac{ \log^*N + \log^* Q + Q\log^*n+ c}{m}.
               \end{align}
$\hfill \Box$

              It is straightforward to plug \eqref{eq:kc_pp} in Corollary \ref{cor:noiseless_normerror} and prove that, for large values of $n$, $O((Q+1)(N+ 2) )$ measurements are sufficient for the successful recovery of the piecewise polynomial functions.

\subsection{Smooth functions} Suppose that $x_1, x_2, \ldots, x_n$ are equispaced samples of  a smooth function $f:[0,1] \rightarrow [0,1]$. 
Let $\mathcal{S}^{\beta}$ represent the class of $\beta+1$ times differentiable functions. For the notational simplicity we assume that $|f^{(m)}(t)| \leq m!$ for every $m \leq \beta+1$. This function is not necessarily a low-complexity signal, but it can be well-approximated by a piecewise polynomial function. To show this, consider partitioning the $[0,1]$ interval into subintervals of size $r_n$, and approximating  the function $f$ with a polynomial of degree $\beta$ in each subinterval. Let $\hat{f}_{\beta}(x)$ denote the resulting piecewise polynomial function. It is straightforward to prove that $\|f-\hat{f}_{\beta}\|_{\infty} \leq r_n^{\beta+1}$. Hence, if $\x_o$ and $\hat{\x}_o$ denote the vectors consisting of the equispaced samples of the original signal and its piecewise polynomial approximation, respectively, it follows that $\|\hat{\x}_o-\x_o\|_2 \leq \sqrt{n} r_n^{\beta+1}$. We can summarize our discussion in the following proposition.
\begin{figure}
\begin{center}
\includegraphics[scale=0.35]{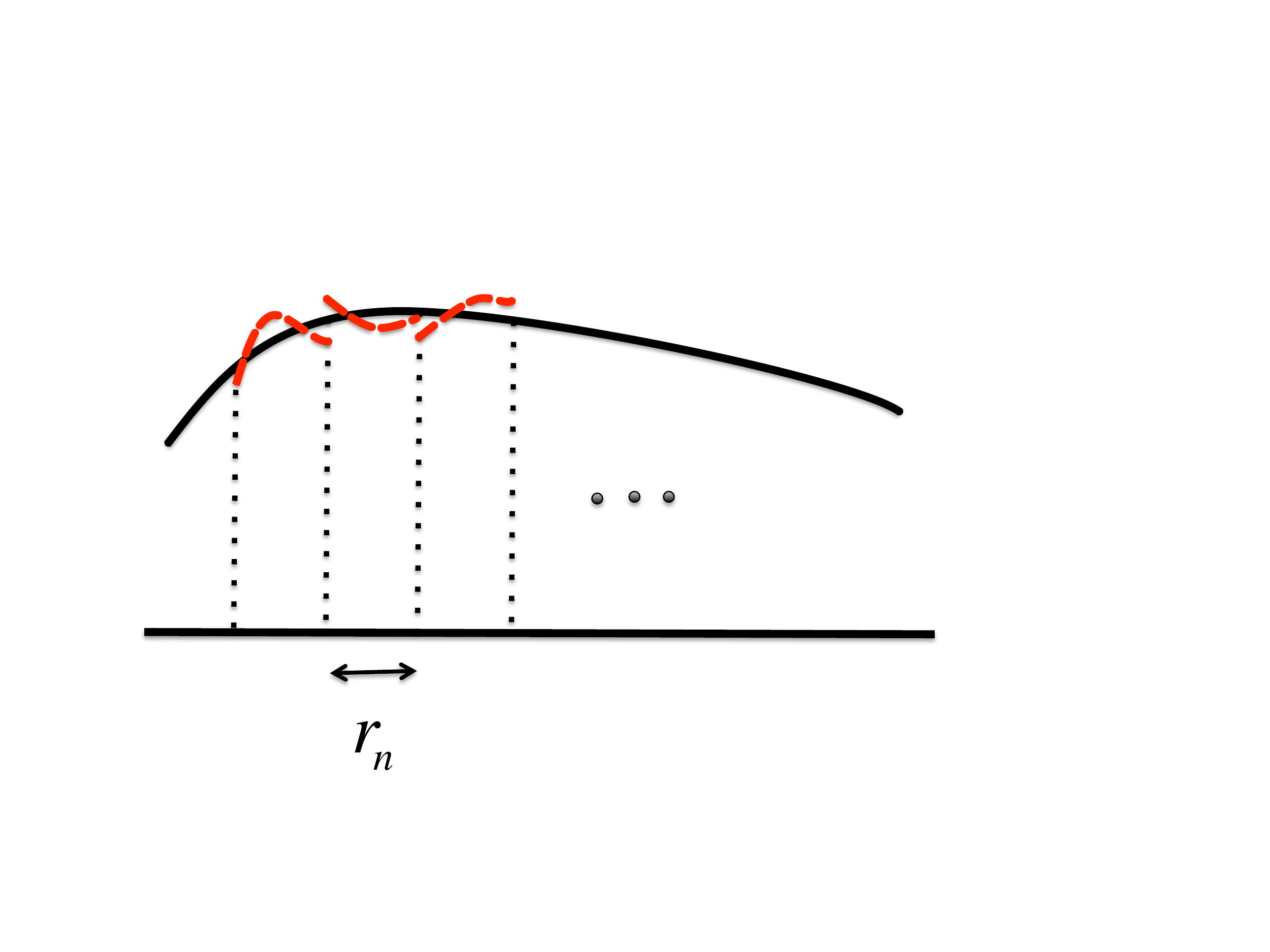}

\vspace{-0.9cm}

\caption{The representation of a smooth function (solid black curve) and its piecewise polynomial approximation (dashed red). As  the subinterval size $r_n$ becomes smaller, the approximation become more accurate.}
\label{fig:smoothfunction}
\end{center}
\end{figure}

\begin{proposition}
For $n\in\mathds{N}$, let $\xv_o\in\mathds{R}^n $ denote the vector of $n$ equispaced samples of   $f \in \mathcal{S}^{\beta}$. Let $\yv=A\xv_o$, where $A$ is a $d\times n$ random matrix with i.i.d.\ $\Nc(0,1)$ entries. Also, let $\xhb$ denote the solution of  low-complexity least square algorithm in \eqref{eq:alg_model_mismatch}, with $m = \log n$ and $\kappa_{m,n} = 2 (2+ \beta) (n^{\frac{2}{2\beta+3}}+1)$. Then, for $n$ large enough and $d = \lceil \kappa_{m,n} \log n \rceil$ and any $\epsilon_1, \e_2>0$, we have
\[
\P(\|\xv_o -\xhb_o\|_2 > {c\over \sqrt{\log n}}) \leq 2\ex^{-0.5d},
\]  
where $c$ is a constant independent of $n$.
\end{proposition}
{\em Proof:}
Partition the $[0,1]$ interval into subintervals of size $r_n = n^{-\frac{1}{\beta+3/2}}$, and  approximate  the function $f$ with a polynomial of degree $\beta$ in each subinterval. Let $\hat{f}_{\beta}$ denote the resulting piecewise polynomial function. According to Proposition \ref{prop:polynomial}, for $n$ sufficiently large, the KID of the samples of $\hat{f}_{\beta}$, ${\bf \xt}_o$, at resolution $m=\lceil\log n\rceil$, is less than $(n^{\frac{1}{\beta+3/2}}+1)(\beta+2)(1+ \delta)$, for any $\delta>0$. Set $\delta=1$ and assume that $n$ is large enough for this result to hold. By Theorem \ref{thm:4}, 
\begin{align*}
\P& \Big(\|\x_{o}-\xhb_{o}\|_2> {1\over \sqrt{1-t}}(\sqrt{n\over d}+2)({1\over{\sqrt{n}}}+2\e_n) +{1\over{\sqrt{n}}}\Big) \nonumber \\
&\leq 2^{ \kappa_{m,n} m} {\rm e}^{\frac{d}{2} (t +\log(1-t) )} + {\rm e}^{- \frac{d}{2} }. \hspace{4cm} 
\end{align*}
Furthermore, as described before,  $\epsilon_n = \|\xtb-\x_o\|_2 \leq  \sqrt{n} r_n^{\beta+1}= n^{-\frac{1}{2}+\frac{1}{2\beta+3}}$. Plugging in $t=0.965$, $d = \lceil \kappa_{m,n} \log n \rceil$, and $\epsilon_n =   n^{-\frac{1}{2}+\frac{1}{2\beta+3}}$ completes the proof.  $\hfill \Box$

\subsection{Low-rank matrices} \label{sec:lowrank}
Let $\mathcal{C}_r(M,N)$ be the class of $M \times N$ real-valued rank-$r$ matrices $X$ with $\sigma_{\rm max}(X) \leq 1$. The following theorem characterizes the KID of a matrix in this class at resolution $m$.

\begin{proposition}\label{prop:low-rank}
Let $X \in \mathcal{C}_r(M,N)$. Then
\[
\kappa_{m,n}(X) \leq   r(M+N+1) + \frac{\log^* r + r(M+N+1)\log(3r) - r+c}{m}.
\]
\end{proposition}
{\em Proof:}
Having access to the values of $M$, $N$ and  the resolution level $m$, consider the program that describes $X$ through its singular value decomposition as follows. Denote the singular value decomposition of the matrix $X$ as $X=U \Sigma V^T$ where $U \in \mathds{R}^{M \times r}$,  $V \in \mathds{R}^{N \times r}$ and $\Sigma \in \mathds{R}^{r \times r}$  is a diagonal matrix. Note that $U^T U = I_r$ and $V^TV =I_r$. To describe $X$, first we use a constant  number of bits to describe the structure of the data  as a  matrix of rank $r$, and also our coding strategy, which is describing the quantized versions of $U$, $\Sigma$, and $V$. To describe the rank $r$, the code uses $\log^* r$ bits. The next step is to describe the quantized versions of $U$, $\Sigma$ and $V$. Let $m_{u}$, $m_v$, and $m_{\sigma}$ denote the resolution levels  used in the  uniform quantization of the elements of $U$, $V$, and $\Sigma$, respectively. Hence,  the quantized matrices can be described using $r M m_{u}+ r N m_v + r m_{\sigma}$ bits. Let $\Uh$, $\Vh$ and $\hat{\Sigma}$ denote the quantized version of $U$, $V$ and $\Sigma$ at the specified resolutions, respectively. Let $\hat{X} \triangleq \hat{U} \hat{\Sigma} \hat{V}$. By the triangle inequality,
\begin{align}
|X_{ij} - \hat{X}_{ij} | &= |\uv_i^T \Sigma \vv_j - \hat{\uv}_i^T \hat{\Sigma} \hat{\vv}_j| \nonumber\\
&\leq  |\uv_i^T \Sigma \vv_j - \hat{\uv}_i^T {\Sigma} {\vv}_j| +  |\hat{\uv}_i^T \Sigma \vv_j - \hat{\uv}_i^T \hat{\Sigma} {\vv}_j| +  |\hat{\uv}_i^T \hat{\Sigma} \vv_j - \hat{\uv}_i^T \hat{\Sigma} \hat{\vv}_j|,
\end{align}  
where $\uv_i^T, \vv_i^T, \hat{\uv}_i^T$, and $\hat{\vv}_i^T$ denote the $i^{\rm th}$ rows of $U$, $V$,  $\hat{U}$ and $\hat{V}$, respectively. Note that $|U_{ij}| \leq 1$, $|V_{ij}|\leq1$, for all $i,j$. Also by assumption,  $\sigma_{\rm max}(\Sigma)\leq1$,  and therefore  $0 \leq \Sigma_{ii} <1$, for $i=1,\ldots,r$. Moreover,  $|U_{ij}-\hat{U}_{ij}|<2^{-m_u+1}$, $|V_{ij}- \hat{V}_{ij}|<2^{-m_v+1}$, and finally $|\Sigma_{ii}- \hat{\Sigma}_{ii}|<2^{-m_{\sigma}}$. Therefore,
\begin{align}
|X_{i,j} - \hat{X}_{i,j} | &\leq  |\uv_i^T \Sigma \vv_j - \hat{\uv}_i^T {\Sigma} {v}_j| +  |\hat{\uv}_i^T \Sigma \vv_j - \hat{\uv}_i^T \hat{\Sigma} {v}_j| +  |\hat{\uv}_i^T \hat{\Sigma} \vv_j - \hat{\uv}_i^T \hat{\Sigma} \hat{\vv}_j| \nonumber \\
&\leq \|\uv_i- \hat{\uv}_i\|_2 \|  \Sigma \vv_j\|_2 + \|\uv_i\|_2 \| (\Sigma- \hat{\Sigma}) \vv_j\|_2 + \|\hat{\uv}_i\|_2 \| \hat{\Sigma} (\vv_j- \hat{\vv}_j)\|_2 \nonumber \\
& \leq \|\uv_i- \hat{\uv}_i\|_2 \sigma_{\rm max}(\Sigma) \|\vv_j\|_2 +   \|\uv_i\|_2  \sigma_{\max}(\Sigma- \hat{\Sigma}) \|\vv_j\|_2 +   \|\hat{\uv}_i\|_2  \sigma_{\max} (\hat{\Sigma})\| (\vv_j- \hat{\vv}_j)\|_2 \nonumber \\
& \leq \sqrt{r 2^{-2m_u+2}} \sqrt{r} + \sqrt{r} 2^{- m_{\sigma}} \sqrt{r}+ \sqrt{r} \sqrt{r 2^{-2m_v+2}}\nonumber\\
& \leq r 2^{-m_u+1} + r2 ^{-m_\sigma} + r2^{-m_v+1}.
\end{align}
To ensure reconstructing the samples at resolution $m$, we have   
\[
 r 2^{-m_u+1} + r2 ^{-m_\sigma} + r2^{-m_v+1} \leq 2^{-m+1}.
\]
Setting $m_u = m_v = m_\sigma+1$, we obtain $m_\sigma \geq m + \log (3r)-1$. Therefore, the KID at resolution $m$ of $X$ is upper bounded as follows:
\begin{align}
\kappa_{m,M,N} &\leq \frac{\log^* r + rM m_u + rN m_v + r m_\sigma +c}{m} \nonumber \\
&\leq \frac{\log^*r+rM(m+ \log(3r)) + rN(m+ \log(3r)) + r(m+ \log(3r)-1) )}{m} \nonumber \\
&\leq r(M+N+1) + \frac{\log^* r + r(M+N+1)\log(3r) - r+c}{m}. \nonumber 
\end{align}
$\hfill \Box$

Consider $m = \lceil \log n \rceil$. If we assume that $M,N, r$ are all sufficiently large while $r \ll M,N$, then we can upper bound $\kappa_{m,M, N} \leq r(M+N+1)(1+ \delta)$, where $\delta$ is small fixed number. It is straightforward to plug this upper bound in Corollary \ref{cor:noiseless_normerror} and prove that, for large values of $M,N$, $3r(M+N+1)(1+ \delta))$ measurements are sufficient for the ``successful'' recovery of the low-rank matrices.


\section{Related work}\label{sec:related}

\subsection{Kolmogorov complexity and applications}
This paper is inspired by \cite{DonohoKS2002} and \cite{DoKaMe06}. \cite{DonohoKS2002} considers the well-studied problem of estimating $\boldsymbol{\theta}\in\mathds{R}^n$ from its noisy observation $\sv = \boldsymbol{\theta} + \zv$, where
$\zv$ represents the noise in the system. It suggests using the \textit{minimum Kolmogorov complexity
estimator} (MKCE)  and proves that if $\{\theta_i\}_{i=1}^n\overset{i.i.d.}{\sim} \pi$, then under several scenarios for the signal
and noise, the average marginal distribution of the estimate derived by MKCE tends to the actual posterior distribution.
\cite{DoKaMe06} considers the problem of CS over real-valued sequences with finite Kolmogorov complexity and defines the Kolmogorov complexity of a real-valued  sequence $\x=(x_1,\ldots,x_n)$ as the length of the program that prints the binary representation of $\x$ and halts.  
Consider the set of all real-valued  sequences with Kolmogorov complexity less than or equal to $k_0$, \ie
\[
\Sc(k_0)\triangleq\{\xv: K(\xv)\leq k_0\}.
\]
Let $A$ denote a $d\times n$ binary matrix, $\xv_o=(x_1,x_2,\ldots,x_n)^T$, $\yv_o=A\xv_o$.  \cite{DoKaMe06} proposes  the following  algorithm for recovering  $\xv_o$ from its linear measurements $\yv_o$:
\begin{align}
{\bf \xh}(\yv_o,A) &\triangleq \argmin_{\yv_o=A\xv} K(\xv).
\end{align} 
It proves that $2k$ random linear  measurements are sufficient for recovering sequences in $\Sc(k_0)$ with high probability. This result does not consider any non-ideality in the signal or the measurements. Furthermore, note that $\Sc(k_0)$  covers none of the classes of signals of interest in CS,  such as sparse vectors or low-rank matrices. Almost all such signals have infinite Kolmogorov complexity, and therefore are not covered by the framework proposed in \cite{DoKaMe06}. Our generalizations require completely different proof techniques. Our paper settles both issues. 

 In  independent work, \cite{BaDu11} and \cite{BaDu12} have explored the performance of an algorithm like  MCP  for CS problems. Replacing the Kolmogorov complexity with the empirical entropy, they propose a Markov chain Monte Carlo approach similar to \cite{JalaliW:08,JalaliW:12, BaWe11} to solve the recovery problem. The empirical results provided in \cite{BaDu12} are very promising. Our theoretical results explain why such algorithms perform well in practice. 

Finally, we should mention that Kolmogorov complexity has proved to be useful in other applications such as similarity detection \cite{CiVi07, Vitanyi12}, density estimation \cite{BaCo91} and compression and denoising \cite{VeVi10}. For more information on the progress in these areas, see \cite{book_vitanyi}.

\subsection{Stochastic models}
In this paper, we considered  deterministic signal models. While deterministic signal models are the most popular models in CS, stochastic models have been also extensively explored; see \cite{HeCa09, Schniter2010, BaGuSh09, RaFlGo10, DoMaMoNSPT, DoMaMo09, DoTa09, WuVe10, MaAnYaBa11, DoJoMaMoEllp, MaDo09sp} and the references therein for more information. The most relevant work, to ours  is \cite{WuVe10}. It considers the problem of recovering a memoryless process from a linear set of measurements and proves a connection between the number of measurements required and the R\'{e}nyi information dimension. The upper information dimension of a random vector $(X_1, X_2, \ldots, X_n)$ is defined as 
\[
\bar{d}(X_1, \ldots, X_n) \triangleq \lim \sup_{m \rightarrow \infty} \frac{H([X_1]_m, \ldots, [X_n]_m)}{m}.
\] 
There is a connection between the KID of a sequence and its R\'{e}nyi information dimension \cite{cover} (Theorem 14.3.1). 
%
In spite of such connections, there are several important differences between our work and the work of \cite{WuVe10}. First, the results in \cite{WuVe10} are asymptotic, and the amount of error and the probability of correct recovery for finite dimensional signals have not been established there. Second, the stochastic approach proposed in \cite{WuVe10} considers a specific distribution that is assumed to be known in the recovery process while we are considering universal schemes in this paper. 

\subsection{Universal schemes and minimum entropy coder}

Our work has some connections with the minimum entropy  decoder proposed by Csiszar in \cite{Csiszar82}. He suggests a universal minimum entropy decoder for reconstructing an i.i.d. signal from its linear measurements at a rate determined by the entropy of the source. For more information, see \cite{CaShVe03, CoMeEf05} and the references therein.

Finally we should emphasize that universal algorithms (that perform ``optimally'' without knowing the distribution of the data) have been explored extensively in information theory and are popular in many applications, including compression \cite{ZiLe78, JalaliW:12}, denoising \cite{WeOrSeVeWe05}, prediction \cite{FeMeGu92}, and many more. However, to the best of our knowledge our results provide the first universal approach for CS.

\subsection{Signal models}
As mentioned in the Introduction, in this paper we have addressed a central problem in the field of CS. Since the early days of CS, there have been many efforts to push the limits of the technique beyond  sparsity. This line of work has resulted in a series of  papers each of which either generalizes the signal model or reduces the required  number of measurements by introducing more structure on the signal; see, for example, \cite{RichModelbasedCS, ChRePaWi10, VeMaBl02, ReFaPa10, ShCh11, HeBa12, HeBa11}. As proved in Section \ref{sec:examp}, some of these models can be considered as subclasses of the general model we consider here.  However, it is worth noting thatm even though the MCP algorithm proposed here is universal, since the Kolmogorov complexity is not computable, it is not immediately useful for practical purposes.


\section{Proofs of the main results} \label{sec:proofs}

\subsection{Useful lemmas} 

The following lemmas are  frequently used in our proofs. 
\vspace{.1cm}

\begin{lemma}[$\chi^2$ concentration]\label{lemma:chi}
Fix $\tau>0$, and let $Z_i\sim\Nc(0,1)$, $i=1,2,\ldots,d$. Then,
\begin{align*}
\P\left( \sum_{i=1}^d  Z_i^2 <d(1- \tau) \right)  \leq {\rm e} ^{\frac{d}{2}(\tau + \log(1- \tau))}
\end{align*}
and
\begin{align}\label{eq:chisq}
\P\left( \sum_{i=1}^d  Z_i^2 > d(1+\tau) \right)  \leq {\rm e} ^{-\frac{d}{2}(\tau - \log(1+ \tau))}.
\end{align}
\end{lemma}
{\em Proof:}
Employing the Chernoff  bound, for any $\lambda>0$, we have
\begin{align}
\P\left( \sum_{i=1}^d  Z_i^2-d < - d \tau \right) &= \P\left(- \sum_i Z_i^2+d > d\tau \right)\nonumber\\
& \leq {\rm e}^{-\lambda d \tau} \E\left[ \rm{e}^{\lambda(d- \sum Z_i^2)} \right] \nonumber \\
 &= {\rm e}^{-\lambda d \tau + \lambda d} \left( \E [{\rm e}^{-\lambda Z_1^2}] \right)^d \nonumber\\
 &= {\rm e}^{-\lambda d \tau + \lambda d} \left(1+ 2\lambda \right)^{-d/2},\label{eq:chisquarupperbound}
\end{align}
where the last line follows from the characteristic function of a Chi-square of degree $d$ \cite{hoggintroduction}. We optimize over $\lambda$ to obtain
\begin{equation}
\lambda^* = \frac{\tau}{2(1- \tau)}. \label{eq:optlambda}
\end{equation}
Plugging \eqref{eq:optlambda} into \eqref{eq:chisquarupperbound}, we obtain \eqref{eq:chisq}.
$\hfill \Box$
\vspace{.1cm}

\begin{lemma}\label{lemma:gaussian-vectors}
Let $\Xv$ and $\Yv$ denote two independent Gaussian vectors of length $n$ with i.i.d.\ elements. Further, assume that for $i=1,\ldots,n$,  $X_i\sim\Nc(0,1)$ and  $Y_i\sim\Nc(0,1)$. Then the distribution of  $\Xv^T\Yv=\sum_{i=1}^nX_iY_i$ is the same as the distribution of $\|\Xv\|_2G$, where $G\sim\Nc(0,1)$ is independent of $\|\Xv\|_2$.
\end{lemma}
{\em Proof:}
Note that
\begin{align}
{\Xv^T\Yv \over \|\Xv\|_2}&=\sum_{i=1}^n{X_i\over \|X^n\|_2}Y_i.
\end{align}
Given $\Xv/\|\Xv\|_2={\bf a}$, 
\[
\sum_{i=1}^n{X_i\over \|X^n\|_2}Y_i\sim \Nc(0,1),
\] 
because $\|\mathbf{a}\|_2^2=1$. Therefore, since the distribution of $\Xv^T\Yv/\|\Xv\|_2$ given $\Xv/\|\Xv\|_2={\bf a}$ is independent of the value of ${\bf a}$,  the unconditional distribution of $\Xv^T\Yv/\|\Xv\|_2$ is also $\Nc(0,1)$.
To prove independence, note that $\Xv/\|\Xv\|_2$ and $\Yv$ are both independent of $\|\Xv\|_2$.
$\hfill \Box$

\vspace{.1cm}

The following lemma is adapted from \cite{Vershyninnotes} (Proposition 5.10). 

\begin{lemma}\label{lem:sub-Gaussiansum}
Let $Z_1, Z_2, \ldots, Z_n$ be i.i.d. zero-mean ${\rm SG}(c_1,c_2)$ random variables. Let $\mathbf{a}= (a_1,a_2, \ldots, a_n) \in \mathds{R}^n$ be a vector satisfying $\|\mathbf{a}\|_2^2=1$. Then 
\[
\P \left(\left|\sum_{i=1}^n a_i Z_i \right| > t \right) \leq c_1 \ex^{-c_2 t^2}.
\] 
In other words $\sum_{i=1}^n a_i Z_i $ is also ${\rm SG}(c_1,c_2)$.
\end{lemma}

\begin{definition}
A random variable $X$ is called {\em subexponential}, denoted by ${\rm SE}(c_1,c_2)$, if and only if 
\[
\P(|X|>t) \leq c_1 \ex^{-c_2t}.
\]
\end{definition}

Slightly modified versions of the proofs we provide in the rest of this section can be found in \cite{Vershyninnotes}. For the sake of clarity and uniformity we state these lemmas with their proofs here.

\vspace{.1cm}

\begin{lemma}\label{lem:subexp_moments}
Let $Z $ be a ${\rm SE}(c_1,c_2)$ random variable. Then, it follows that
\begin{align*}
\E[|Z|^p] &\leq \frac{2c_1 p!}{c_2^p}.
\end{align*}
\end{lemma}
{\em Proof:}
Here we prove this lemma for the case where $p$ is even. The other case follows the same approach. Let $F(z)$ denote the cumulative distribution function of the random variable $Z$  
 \begin{align*}
\E[|Z|^p] &= \int_0^{\infty} z^p dF(z) + \int_{-\infty}^{0}z^p dF(z) \overset{(a)}{=} \int_{0}^{\infty} p z^{p-1} \int_{z}^{\infty} dF(x) dz - \int_{-\infty}^{0} p z^{p-1} \int_{-\infty}^{z} dF(x)dz \nonumber \\
& \leq \int_{0}^{\infty} p z^{p-1} c_1 \ex^{-c_2 z}dz - \int_{-\infty}^{0} p z^{p-1} c_1 \ex^{c_2 z}dz = \frac{2c_1 (p !)}{c_2^p}.
\end{align*}
Equality (a) is the result of integration by parts. 
$\hfill \Box$

\begin{lemma} \label{lem:exponentialexpo}
Let $Z$ be a zero-mean ${\rm SE}(c_1,c_2)$ random variable. Then we have
\begin{align*}
\E\left[\ex^{\lambda Z}\right] &\leq \ex^{4c_1\lambda^2/c_2^2}, \ \ \ \ \forall \lambda < c_2/2.
\end{align*}
\end{lemma}

{\em Proof:}
We prove this theorem by expanding the exponential function $\ex^{\lambda Z}$ and bounding the moments using Lemma \ref{lem:subexp_moments} as follows:
\begin{align}
\E\left[\ex^{\lambda Z}\right] &= \E \left[1+ X + \sum_{k=2}^{\infty} \frac{\lambda ^k X^k}{K!}\right] =  \E \left(1+ \sum_{k=2}^{\infty} \frac{\lambda ^k X^k}{K!} \right) \nonumber \\
&\leq 1+ 2c_1\left( \left(\frac{\lambda}{c_2}\right)^2 + \left(\frac{\lambda}{c_2}\right)^3+ \ldots \right) \leq 1+ 2c_1 \left(\frac{\lambda}{c_2}\right)^2 \left(\frac{1}{1- \lambda/c_2}\right).
\end{align}
Assuming that $\frac{\lambda}{c_2}<\frac{1}{2}$, we obtain
\[
\E\left[\ex^{\lambda Z}\right]  \leq 1+4 c_1 \left(\frac{\lambda}{c_2}\right)^2 \leq \ex^{4c_1\lambda^2/c_2^2},
\]
where the last inequality is due to the fact that $1+x \leq \ex^x$ for $x\geq 0$.
$\hfill \Box$

\vspace{.1cm}

\begin{lemma}\label{lem:subexponentialsquare}
Let $Z_1, Z_2, \ldots, Z_n$ be i.i.d.\ ${\rm SG}(c_1,c_2)$ random variables with mean zero and variance 1. Then we have
\[
\P\left( \left|\sum_{i=1}^n ( Z^2_i-1)\right| > nt\right) \leq 2 \ex^{-nc_2^2 t^2/16c_3},     \ \ \ {\rm for} \ t \in (0, \frac{c_3}{c_2}),
\] 
where $c_3 \triangleq \max(\ex^{c_2}, c_1 \ex^{-c_2})$.
\end{lemma}

{\em Proof:}
Define $X_i \triangleq Z_i^2-1$. It is straightforward to confirm that for all $t>1$,
\begin{equation}\label{eq:upperexpon}
\P(|X_i| >t ) \leq c_1 \ex^{-c_2(t+1)}.   
\end{equation}
Define $c_3 \triangleq \max(\ex^{c_2}, c_1 \ex^{c_2})$. If we combine the fact that $\P(|X_i|>t) \leq 1$ for $0\leq t \leq 1$ with \eqref{eq:upperexpon}, we obtain
\[
\P(|X_i|>t) \leq c_3 \ex^{-c_2t}.
\]
We have
\begin{align}\label{eq:markovupper}
\P\Big(\sum_i X_i >nt \Big) = \P\Big(\ex^{\lambda \sum_i X_i} > \ex^{\lambda nt} \Big) \leq \ex^{-\lambda nt } \left( \E\left[\ex^{\lambda X_1}\right]\right)^n \leq \ex^{-\lambda n t +4n c_3 \lambda^2/c_2^2},
\end{align}
where the last inequality is the result of Lemma \ref{lem:exponentialexpo}. Assuming $t < \frac{c_3}{c_2}$ and setting $\lambda = tc_2^2 /(8c_3)$, we obtain
\[
\P\left(\sum_{i=1}^n X_i >nt \right)  \leq \ex^{\frac{-n(c_2t)^2}{16c_3}}.
\]
Using the same argument, we find a similar upper bound for $\P(\sum_{i=1}^n X_i <-nt)$. 
$\hfill \Box$

\vspace{.1cm}

\begin{lemma} \label{lem:subgaussspectrum}
Let $A$ be a $d \times n$ matrix with i.i.d.\ ${\rm SG}(c_1,c_2)$ elements, and suppose that the elements satisfy $\E(A_{ij})=0$ and $\E(A_{ij}^2)=1$. Then there exist two constants $c'_1,c'_2$ depending only on $c_1$ and $c_2$ such that with probability at least $1- \ex^{-c'_2 t^2}$,
\[
\sigma_{\rm max}(A) \leq \sqrt{d} + c'_1\sqrt{n}+t.
\]
\end{lemma}
{\em Proof:}
See Theorem 5.39  in \cite{Vershyninnotes} for more information on the proof and the constants that are involved.
$\hfill \Box$

\subsection{Proof of Theorem \ref{thm:1}} \label{sec:proofthm1}
Let $\xhb_o$ denote the solution of MCP, and let  $\qhb_m \triangleq \xhb_o-\phi_m(\xhb_o)$ denote the quantization error of the reconstructed signal at resolution $m$, where for $\xv\in[0,1]^n$, $\phi_m(\xv)$ is defined in Remark \ref{remark:1}. 

Since both $A\xv_o =\yv_o$ and $A\xhb_o=\yv_o$, it follows that 
\begin{align}
A\xv_o&=A(\phi_m(\xhb_o)+\qhb_m)\nonumber
\end{align}
and 
\begin{align}
A(\xv_o-\phi_m(\xhb_o))&=A\qhb_m.
\end{align}
On the other hand, by definition, $\|\qhb_m\|_{\infty}\leq 2^{-m}$, and therefore
\[
\|\qhb_m\|_2\leq 2^{-m}\sqrt{n}.
\] 
Hence,
\begin{align}
\|A(\xv_o- \phi_m(\xhb_o))\|_2&=\|A\qhb_m\|_2 \nonumber\\
&\leq \sigma_{\max}(A) 2^{-m}\sqrt{n},\label{eq:upper_bd}
\end{align}
\noindent where $\sigma_{\rm max}(A)$ is the maximum singular value of matrix $A$. By definition, $K^{[\cdot]_{m}}(\xv_o) \leq \kappa_{m,n}m$, and since $\xhb_o$ is the solution of \eqref{eq:alg}, we have
\begin{align}
 K^{[\cdot]_{m}}(\xhb_{o}) \leq K^{[\cdot]_{m}}(\x_{o})\label{eq:kol_xho} \leq \kappa_{m,n}m.
\end{align}
Define  set $\mathcal{S}$ as
\[
\Sc\triangleq\left\{\xv_o-\phi_m(\xtb_o): \; \xtb_o \in[0,1]^n,\,     K(\phi_m(\xtb_o)) \leq \kappa_{m,n}m \right\}.
\]
Define  event $\Ec_1^{(n)}$ as
\begin{align}
\Ec_1^{(n)}\triangleq\{ \forall \ \mathbf{h} \in \Sc \, :  \,  \|A \mathbf{h}\|_2 >  \sqrt{d(1-t)} \|\mathbf{h}\|_2 \},\label{eq:E1}
\end{align}
and,   event  $\Ec_2^{(n)}$ as
\begin{align}
\Ec_2^{(n)}\triangleq \left\{\sigma_{max}(A) - \sqrt{d} - \sqrt{n} < \sqrt{d} \right\}.\label{eq:E2}
\end{align}

Conditioned on $\Ec_1^{(n)}\cap \Ec_2^{(n)}$, we have
\begin{align}\label{eq:pf_thm2_bound}
\|\xv_o  - \xhb_o\|_2 &= \left\|\xv_o- \phi_m(\xhb_o)-\qhb_m\right\|_2 \nonumber\\
& \leq\left\|\xv_o - \phi_m(\xhb_o)\|_2  + \| \qhb_m\right\|_2 \nonumber\\
& \overset{(a)}{ \leq} {\| A(\xv_o - \phi_m(\xhb_o)) \|_2 \over \sqrt{d(1-t)}}+ 2^{-m}\sqrt{n} \nonumber\\
&\overset{(b)}{ \leq} {\sigma_{\max}(A)2^{-m}\sqrt{n} \over  \sqrt{d(1-t)}}+2^{-m}\sqrt{n}\nonumber\\
&\overset{(c)}{ \leq} ({\sqrt{n}+2\sqrt{d}\over \sqrt{d(1-t)}}+1)2^{-m}\sqrt{n}\nonumber\\
&{\leq} \left((1-t)^{-0.5}\left(\sqrt{n/d}+2\right) +1\right)2^{-m}\sqrt{n}.
\end{align}
Inequality (a) holds since due to $\Ec_1^{(n)}$,  $ \| A(\xv_o  - \phi_m(\xhb_o)) \|_2 \geq \sqrt{(1-t)d} \| (\xv_o - \phi_m(\xhb_o)) \|_2$. Inequality (b) is a result of \eqref{eq:upper_bd}, and inequality (c) is due to  $\Ec_2^{(n)}$. Hence, 
\begin{align}\label{eq:pconditional}
\P\left(\| \xv_o -\xhb_o\|_2 >\e,\Ec_1^{(n)} \cap\Ec_2^{(n)}\right)=0,
\end{align}
where $\e\triangleq ((1-t)^{-0.5}(\sqrt{nd^{-1}}+2) +1)2^{-m}\sqrt{n}$. Using these definitions and the union bound, we have
\begin{align}\label{eq:proberror}
\P\left(\| \xv_o -\xhb_o\|_2 >\e\right) & = \P\left(\| \xv_o -\xhb_o\|_2 >\e,\Ec_1^{(n)} \cap\Ec_2^{(n)}\right) + \P\left(\| \xv_o -\xhb_o\|_2 >\e,\Ec_1^{(n), c} \cup\Ec_2^{(n), c}\right)=\nonumber \\
 & = \P\left(\| \xv_o -\xhb_o\|_2 >\e \ | \ \Ec_1^{(n), c} \cup\Ec_2^{(n), c}\right) \P \left(\Ec_1^{(n), c} \cup\Ec_2^{(n), c}\right) \nonumber\\
& \leq  \P\left(\Ec_1^{(n),c}\right)+\P\left(\Ec_2^{(n),c}\right).
\end{align}

On the other hand, by Lemma \ref{lemma:chi}, for fixed  $\x\in\mathds{R}^n$,
\begin{align*}
\P\left(\|A\x\|_2 \leq  \sqrt{(1-t)d} \|\x\|_2\right) &=\P\left(\left\|A\frac{\x}{\|\x\|_2}\right\|_2^2 \leq (1-t)  d \right)\nonumber\\
&= \P\left(\sum_{i=1}^d Z_i^2 \leq (1-t) d \right)\nonumber\\
& \leq \ex^{\frac{d}{2}(t+  \log (1-t)) },
\end{align*}
where, for $i=1,\ldots,d$, $Z_i\triangleq\|\x\|^{-1}_2\sum_{j=1}^{n} A_{i,j}x_{j}$. Therefore, since $|\Sc|\leq 2^{\kappa_{m,n}n}$, by the union bound, it follows that
\begin{align}\label{eq:event1_nless}
&\P\left(\Ec_1^{(n),c}\right) \leq 2^{\kappa_{m,n}m}  \ex^{\frac{d}{2}(t+  \log (1-t)) }.
\end{align}
Finally, using the results on the  concentration of Lipschitz functions of a Gaussian random vector \cite{CaTa05}, 
\begin{align}\label{eq:event2_nless}
\P\left(\Ec_2^{(n),c}\right) &= \P\left(\sigma_{max}(A) - \sqrt{d} - \sqrt{n} >  \sqrt{d} \right)\nonumber\\ 
&\leq {\rm e}^{-d/2}.
\end{align}
Plugging \eqref{eq:pconditional}, \eqref{eq:event1_nless}, and \eqref{eq:event2_nless} into \eqref{eq:proberror} completes the proof.
$\hfill \Box$

\subsection{Proof of Theorem \ref{thm:noisysetting}} \label{sec:proofthmnoisy}
Remember that $\xhb_o=\xh_o^n$ denotes the solution of 
\begin{align}
\min &\;\;\; \;\; \|A\x-\yv_o\|_2, \nonumber \\
{\rm s.t.} &\;\;\;\;\; K^{[\cdot]_{m_n}}(\x) \leq \kappa_n m.\label{eq:alg_noisy}
\end{align} 
By the assumption of the theorem, $K^{[\cdot]_m}(\xv_o ) \leq k_{n} m$. Therefore, $\xv_o $ is  a feasible point in \eqref{eq:alg_noisy}, and we have
\begin{align}
\|A\xhb_o-\yv_o\|^2_2&\leq \|A \xv_o -\yv_o\|^2_2\nonumber\\
&=\|A \xv_o -A\xv_o -\w\|^2_2=\|\w\|^2_2.\label{eq:basic-ineq}
\end{align}
Expanding $\|A\xhb_o-\yv_o\|^2_2=\|A\xhb_o-A\xv_o -\w\|^2_2$ in \eqref{eq:basic-ineq}, it follows that 
\begin{align}
&\|A(\xhb_o-\xv_o )\|^2_2 + \|\w\|^2_2  -2\w^TA(\xhb_o-\xv_o ) \leq \|\w\|^2_2.\label{eq:basic-ineq-expanded}
\end{align}
Canceling $ \|\w\|^2_2$ from both sides of \eqref{eq:basic-ineq-expanded}, we obtain
\begin{align*}
\|A(\xhb_o-\xv_o )\|^2_2 &\leq   2\w^TA(\xhb_o-\xv_o )
\leq   2\left|\w^TA(\xhb_o-\xv_o )\right|.
\end{align*}

Let  $\qhb_m\triangleq\xhb_o-\phi_m(\xhb_o)$, where $\phi_m(\cdot)$ is defined in \eqref{eq:def-of-phi-m}. Using this definition and the Cauchy-Schwartz inequality, we derive a lower bound on $\|A(\xhb_o-\xv_o )\|^2_2$ as 
\begin{align}
\|A&(\xhb_o-\xv_o )\|^2_2 \nonumber\\
&= \|A(\phi_m(\xhb_o)+\qhb_m - \xv_o)\|^2_2 \nonumber\\
 &= \|A(\phi_m(\xhb_o)- \xv_o )+A\qhb_m\|^2_2 \nonumber\\
&\geq \|A(\phi_m(\xhb_o)- \xv_o )\|_2^2
-2\left| \qhb_m ^TA^TA\left(\phi_m(\xhb_o)- \xv_o \right)\right|\nonumber\\
&\geq \|A(\phi_m(\xhb_o)- \xv_o)\|_2^2 -2\left\|A\qhb_m \right\|_2\left\|A\left(\phi_m(\xhb_o)- \xv_o\right)\right\|_2. \label{eq:lower}
\end{align}
On the other hand, again using our definitions plus the Cauchy-Schwartz  inequality, we find an upper bound on $|\w^TA(\xhb_o-\xv_o )|$ as 
\begin{align}
\left|\w^TA(\xhb_o-\xv_o )\right|&
=\left|(\phi_m(\xhb_o) - \xv_o +\qhb_m)^TA^T\w\right|\nonumber\\
&\leq \left|(\phi_m(\xhb_o) - \xv_o )^TA^T\w\right|+\left|\qhb_m^TA^T\w\right|\nonumber\\
&\leq \left|( \phi_m(\xhb_o) - \xv_o)^TA^T\w\right|+\|\qhb_m\|_2\|A^T\w\|_2.\label{eq:upper}
\end{align}
By definition, $\|\qhb_m\|_{\infty}\leq 2^{-m}$. Therefore, 
\begin{align}
\|\qhb_m\|_2\leq2^{-m} \sqrt{n}.\label{eq:ell2-error}
\end{align}
Define $\Delta\triangleq\| \phi_m(\xhb_o)- \xv_o\|_2$, and
\[
\uv\triangleq {A(\phi_m(\xhb_o)- \xv_o)\over \Delta}.
\]
By this definition, combining  \eqref{eq:lower} and \eqref{eq:upper} yields
\begin{align}
 \|\uv\|_2^2\Delta^2  \leq 2(\left\|A\qhb_m \right\|_2\|\uv\|_2 + \left|\w^T\uv\right|)\Delta+2\|\qhb_m\|_2\|A^T\w\|_2.
 \end{align}
For $t_1,t_2,t_3,t_4,t_5>0$, define   events $\Ec_1^{(n)},\ldots,\Ec_5^{(n)}$ as 
\[
\Ec_1^{(n)}\triangleq \{\|\uv\|^2_2 \geq  d(1-t_1) \},
\]
\[
\Ec_2^{(n)}\triangleq \{\|\uv\|^2_2 \leq d(1+t_2) \},
\]
\[
\Ec_3^{(n)}\triangleq \{|\w^T \uv|\leq \sigma\sqrt{(1+t_3)d}  \},
\]
\[
\Ec_4^{(n)}\triangleq \left\{\sigma_{\max}(A)   <\sqrt{d}+\sqrt{n}+t_4 \right\},
\]
and
\[
\Ec_5^{(n)}\triangleq\{\|A^T\mathbf{w}\|_2^2 \leq nd(1+t_5)\sigma^2  \}.
\]

First, we find an upper bound on  $\P((\Ec_1^{(n)}\cap\ldots\cap\Ec_5^{(n)})^c)$.

Define the set $\Sc$ as follows
\[
\Sc\triangleq\left\{\phi_m(\xtb_o)- \xv_o: \; \xtb_o \in[0,1]^n,\,     K(\phi_m(\xtb_o)) \leq \kappa_{n}m \right\}.
\]
Note that $|\Sc|\leq 2^{\kappa_{n}m }$. Given $\phi_m(\xtb_o)- \xv_o\in\Sc$, $A(\phi_m(\xtb_o)- \xv_o)/\|\phi_m(\xtb_o)- \xv_o\|_2$ is a vector of length $d$ with i.i.d. entries distributed as $\Nc(0,1)$.  Therefore, by   Lemma \ref{lemma:chi} and the union bound, we obtain
\begin{align}
\P(\Ec_1^{(n),c})\leq 2^{\kappa_{n}m }{\rm e}^{{d\over 2} (t_1+\log(1-t_1))},\label{eq:E1}
\end{align}
and
\begin{align}
\P(\Ec_2^{(n),c})\leq 2^{\kappa_{n}m }{\rm e}^{-{d\over 2}(t_2-\log(1+t_2))}.\label{eq:E2}
\end{align}
To bound $\P(\Ec_3^{(n),c})$, for $\phi_m(\xtb_o)- \xv_o\in\Sc$,  let $\tilde{\uv}\triangleq {A(\phi_m(\xtb_o)- \xv_o)\over \|\phi_m(\xtb_o)- \xv_o\|_2}$. By Lemma \ref{lemma:gaussian-vectors}, $\w^T\tilde{\uv}$ is distributed as $ \|\w\|_2G$, where $G\sim\Nc(0,1)$ and is independent of $\|\w\|_2$. Therefore,
\begin{align}
\P(|\w^T \tilde{\uv}|\geq \sigma\sqrt{(1+t_3)d} ) &=\P\left(|\w^T \tilde{\uv}|\geq \sigma\sqrt{(1+t_3)d},  \|\w\|_2\geq \sigma\sqrt{(1+\tau)d} \right)\nonumber\\
&\;\;\;\; +\P\left(|\w^T \tilde{\uv}|\geq \sigma\sqrt{(1+t_3)d},  \|\w\|_2< \sigma\sqrt{(1+\tau)d}\right)\nonumber\\
&\leq\P\left( \|\w\|_2\geq \sigma\sqrt{(1+\tau)d} \right)\nonumber\\
&\quad+\P\left(\left\|\w\right\|_2 G\geq  \sigma\sqrt{(1+t_3)d}\left| \|\w\|_2< \sigma\sqrt{(1+\tau)d}\right. \right)\nonumber\\
&\leq \P\left(\|\w\|_2\geq \sigma\sqrt{(1+\tau)d} \right)+\P\left(G\geq \sqrt{1+t_3 \over  1+\tau}\right)\nonumber\\
&\leq \ex^{-{d\over 2}(\tau-\log(1+\tau))}+\ex^{-{1+t_3\over 2(1+\tau)}}.\label{eq:cond-w}
\end{align}
Hence, by the union bound, and the fact that  $|\Sc|\leq 2^{\kappa_{n}m}$, we obtain 
\begin{align} 
\P(\Ec_3^{(n),c}) \leq  2^{\kappa_{n}m}\left( \ex^{-{d\over 2}(\tau-\log(1+\tau))}+\ex^{-{1+t_3\over 2(1+\tau)}}\right).\label{eq:E3}
\end{align}
For $\Ec_4$, it can be shown that \cite{CaTa05}
\begin{align}
\P\left(\Ec_4^{(n),c}\right) =\P\left(\sigma_{\max}(A) < \sqrt{d}+\sqrt{n}+t_4\right) \leq {\rm e}^{- t_4^2/2}. \label{eq:E4}
\end{align}
Finally, to bound $\Ec_5^{(n),c}$, note that given $\w$, $A^T\w$ is an $n$-dimensional i.i.d.~zero-mean variance $\|\w\|_2^2$ normal vector. Therefore, similar to the derivation of  \eqref{eq:cond-w}, we have
\begin{align}
\P(\Ec_5^{(n),c})&=\P\left(\|A^T\w\|_2^2 \geq nd(1+t_5)\sigma^2 \right)\nonumber\\
&\leq \P\left(\|A^T\w\|_2^2 \geq nd(1+t_5)\sigma^2 \left| \|\w\|_2^2\leq d\sigma^2(1+\tau')\right. \right)\nonumber\\
&\;\;+\P\left( \|\w\|_2^2\geq d\sigma^2(1+\tau') \right)\nonumber\\
&\leq   {\rm e} ^{-\frac{n}{2}(t_5 - \log(1+ t_5))} +  {\rm e} ^{-\frac{d}{2}(\tau' - \log(1+ \tau'))},\label{eq:E5}
\end{align}
where $t_6>0$ satisfies $1+t_6= (1+t_5)/(1+\tau')$.

Choosing $t_1=0.5$, from \eqref{eq:E1} and the fact that $d=8r\kappa_{n}m$, which yields $\kappa_nm\leq d/8$, we obtain
\[
 \P(\Ec_1^{(n),c}) \leq {\rm e}^{d(\log 2/8+ 0.5 (0.5+\log 0.5))} \leq \ex^{-0.01d}.
\]
For  $t_2=1.25$, since again $\kappa_nm\leq d/8$, 
\[
 \P(\Ec_2^{(n),c}) \leq {\rm e}^{d(\log 2/8-0.5 (1.25+\log 1.25))} < \ex^{-0.1d}.
\]
For $\tau=1$, and $t_3=4m\kappa_{n}-1$,  from \eqref{eq:E3}, we obtain
\begin{align*}
\P(\Ec_3^{(n),c})& \leq  2^{\kappa_{n}m}\Big( \ex^{-{d\over 2}(1-\log 2)}+\ex^{-m\kappa_n}\Big)\\
&< \ex^{-0.06d}+ \ex^{-0.3m\kappa_n}.
\end{align*}
Choosing $t_4=\sqrt{d}$, from \eqref{eq:E4},  $\P\left(\Ec_4^{(n),c}\right) < \ex^{-0.5d}$. Finally, setting $\tau'=1$ and $t_5=3$, \eqref{eq:E5} yields
\[
 \P(\Ec_5^{(n),c}) \leq  {\rm e} ^{-\frac{n}{2}(3 - \log4)} +  {\rm e} ^{-\frac{d}{2}(1- \log2)} <\ex^{-0.8n}+\ex^{-0.15d}.
\]
Therefore, combining all the bounds, it follows that 
\begin{align}
&\P\left((\Ec^{(n)}_1\cap \Ec^{(n)}_2\cap\Ec^{(n)}_3\cap\Ec^{(n)}_4\cap\Ec^{(n)}_5)^c\right)\nonumber\\
&< 6\ex^{-0.01d}+ \ex^{-0.3m\kappa_n}.
\end{align}


On the other hand,  conditioned on $\Ec_1^{(n)}\cap\ldots\cap\Ec_5^{(n)}$,  we have
\begin{align*}
(1-t_1)d\Delta^2-2\Delta (2^{-m}\sqrt{n}(\sqrt{d}+\sqrt{n}+t_4)\sqrt{d(1+t_2)}+\sigma \sqrt{(1+t_3)d})-2\sigma2^{-m}n\sqrt{(1+t_5)d}\leq 0,
\end{align*}
or, inserting the values of $t_1,\ldots,t_5$ and noting that $d=8rm\kappa_n$, 
\begin{align}
\Delta^2-2\Delta ({6 \over \sqrt{n}}+{3 \over \sqrt{d}}+ \sigma\sqrt{2\over r}\;)-{8\sigma\over \sqrt{d}}\leq 0,\label{eq:main-2nd-order-ineq}
\end{align}
where we have also used the fact that for $m=\lceil\log n\rceil$, $n2^{-m}\leq 1$.  Inequality \eqref{eq:main-2nd-order-ineq} involves a quadratic function in $\Delta$, which has a positive root and a negative root. Hence, for \eqref{eq:main-2nd-order-ineq} to hold, we need $\Delta$ to be smaller than its positive root, which yields
\[
\Delta \leq ({6 \over \sqrt{n}}+{3 \over \sqrt{d}}+ \sigma\sqrt{2\over r}\;) +\sqrt{({6 \over \sqrt{n}}+{3 \over \sqrt{d}}+ \sigma\sqrt{2\over r}\;)^2+{8\sigma\over \sqrt{d}}}.
\]
Finally,
\begin{align}
\|\xv_o-\xhb_o\|_2&\leq \|\xv_o-\phi_m(\xhb_o)\|_2+\|\phi_m(\xhb_o)- \xhb_o\|_2\nonumber\\
&\leq \Delta+\sqrt{n} 2^{-m}.
\end{align}
Therefore, for $n$ and $d$ large enough,
\[
\|\xv_o-\xhb_o\|_2 \leq {3\sigma \over \sqrt{r}}.
\]
This completes the proof of Theorem \ref{thm:noisysetting}. 
$\hfill \Box$


\subsection{Proof of Theorem \ref{thm:4}} \label{sec:proof2}
Since the proof of this theorem is similar to the proof of Theorem \ref{thm:1}, we skip most of the steps and only emphasize the main differences. 
Let $\xhb_o$ denote the solution of \eqref{eq:alg_model_mismatch}.
Define  $\qhb_m$ as the quantization error  $\xhb_o$, i.e.,  $\qhb_m \triangleq \xhb_o- \phi_m(\xhb_o)$. Since $ \|A\xtb_o -\y_o\|_2 =\| A(\xtb_o-\xv_o ) \|_2 \leq  \sigma_{max}(A) \epsilon_n$ and $\xhb_o$ is the minimizer  of \eqref{eq:alg_model_mismatch}, it follows that $\|A\xhb_o-\y_o \| \leq  \sigma_{max}(A) \epsilon_n$. Therefore,
\begin{align}
\|A\xtb_o - A\xhb_o \|_2&=\|A\xtb_o-\y_o - (A\xhb_o-\y_o) \|_2 \nonumber\\
&\leq 2\sigma_{\max}(A)\e_n.\label{eq:normeq1}
\end{align}
Again, by the triangle inequality, 
\begin{eqnarray}
\lefteqn{\|A\xtb_o - A\xhb_o \|_2} \nonumber\\
&=&\|A \xtb_o- A( \phi_m(\xhb_o)+\qhb_m) \|_2 \nonumber\\
& \geq& \|A(\xtb_o- \phi_m(\xhb_o))\|_2 -  \|A\qhb_m\|_2\nonumber\\
& \geq& \|A(\xtb_o- \phi_m(\xhb_o))\|_2 - \sigma_{max}(A) \|\qhb_m\|_2\nonumber\\
& \geq& \|A(\xtb_o- \phi_m(\xhb_o))\|_2 - \sigma_{max}(A) 2^{-m}\sqrt{n}.\label{eq:normeq2}
\end{eqnarray}
Combining \eqref{eq:normeq1} and \eqref{eq:normeq2}, it follows that
\begin{align}
 \|A(\xtb_o- \phi_m(\xhb_o))\|_2 \leq \sigma_{max}(A) 2^{-m}\sqrt{n}+2\sigma_{\max}(A)\e_n.
\end{align}
We also have: $K^{[\cdot]_m}(\xhb_o)  \leq  m \kappa_{m,n}$ and $ K^{[\cdot]_m}(\xtb_o) \leq m \kappa_{m,n}$.

Define the events $\Ec^{(n)}_1$ and $\Ec^{(n)}_2$ as done in \eqref{eq:E1} and \eqref{eq:E2} in the proof of Theorem \ref{thm:1}. Then, applying the  argument used there, it follows that
\begin{align}
\P\left(\| \xv_o -\xhb_o\|_2 >\e\right)
\leq &\P\left(\| \xv_o -\xhb_o\|_2 >\e,\Ec_1^{(n)} \cap\Ec_2^{(n)}\right)\nonumber\\
&+\P\left(\Ec_1^{(n),c}\right)+\P\left(\Ec_2^{(n),c}\right).
\end{align}
The rest of the proof is exactly the same as that for Theorem \ref{thm:1}.
$\hfill \Box$

\subsection{Proof of Theorem \ref{thm:sub-Gaussian}} \label{sec:proofsubgauss}
Let $\xhb_o$ be the solution of the MCP algorithm and $\q_m \triangleq \xv_o - \phi_m(\xv_o)$ and $\qhb_m \triangleq \xhb_o- \phi_m(\xhb_o)$ denote the quantization errors of the original and the reconstructed signals at resolution $m$, respectively. Following exactly the same steps as the proof of Theorem \ref{thm:1}, we obtain
\begin{align}
K^{[\cdot]_m} (\xhb_o) \leq K^{[\cdot]_m} (\xv_o ) \leq \kappa_{m,n} m
\end{align}
and
\begin{align}
\|A(\phi_m(\xv_o)- \phi_m(\xhb_o)) = \sigma_{\rm max}(A) \sqrt{n2^{-2m+2}}.
\end{align}
Since we are dealing with subgaussian random matrices, we define slightly different events here. Let the set $\mathcal{S}_o$ as
\[
\Sc_o\triangleq\left\{\mathbf{h} \, : \, \mathbf{h}= \phi_m(\xhb_o)- \phi_m(\xv_o), \, \xhb_o,\xv_o\in[0,1]^n,\,   K(\phi_m(\xhb_o)) \leq \kappa_{m,n}m, \,   K(\phi_m(\xv_o)) \leq \kappa_{m,n}m \right\},
\]
and define
\begin{eqnarray}
\Ec_1^{(n)} &\triangleq&\{ \nexists \ \mathbf{h} \in \Sc_o   \, ;  \|A (\mathbf{h})\|_2 < \tau \sqrt{d} \|\mathbf{h}\|_2 \}, \ \ \ \ \label{eq:E1-sg} \\
\Ec_2^{(n)}&\triangleq& \left\{\sigma_{max}(A)  <  \sqrt{d} + (c'_2+1)\sqrt{n} \right\},\label{eq:E2-sg}
\end{eqnarray}
where $c'_2$ is the constant introduced in Lemma \ref{lem:subgaussspectrum}. $\P(\|\xv_o - \xhb_o\|> \epsilon)$ can be upper bounded by
\[
\P(\|\xv_o - \xhb_o\|> \epsilon) \leq \P(\| \xv_o  - \xhb_o\|_2> \epsilon, \Ec_1^{(n)} \cap \Ec_2^{(n)}) + \P(\Ec_1^{(n)}) + \P(\Ec_2^{(n)}).
\]
If $A \in \Ec_1^{(n)} \cap \Ec_2^{(n)}$, then similar to \eqref{eq:pf_thm2_bound} we can prove
\[
\|\xv_o  - \xhb_o\|_2 \leq \left(\tau^{-1} \left(\sqrt{(c'_2+1)nd^{-1}+1}\right)+1 \right)\sqrt{n2^{-2m+2}}.
\]
Hence, 
\[
\P(\| \xv_o -\xhb_o\|_2 > \epsilon, \Ec_1^{(n)} \cap \Ec_2^{(n)}) =0.
\]
Also, according to Lemma \ref{lem:subgaussspectrum}, $\P(\Ec_2^{(n)}) \leq \ex^{-c'_1 n}$. Therefore, the main difference is in the calculation of $\P(\Ec_1^{(n)})$:
\begin{align*}
\P\left(\|A \x\|_2 \leq \tau \sqrt{d} \| \x\|_2\right) &=\P\left(\left\|A\frac{\x}{\| \x\|_2}\right\|_2^2 \leq \tau^2  d \right)\nonumber\\
&= \P\left(\sum_{i=1}^d Z_i^2 \leq \tau^2 d \right),
\end{align*}
where for $i = 1, 2, \ldots, d$, $Z_i = \|\x\|_2^{-1}\sum_j A_{ij} x_j$. Therefore, by Lemma \ref{lem:sub-Gaussiansum} we obtain
\[
P(|Z_i|> t) \leq c_1 \ex^{-c_2 t}. 
\] 
According to Lemma \ref{lem:subexponentialsquare} we have 
\[
\P\left(\sum_{i=1}^d Z_i^2 \leq \tau^2 d \right)< \ex^{-\frac{dc_2^2(\tau^2-1)^2}{16c_3} },
\]
where $c_3 \triangleq \max (c_1 \ex^{-c_2}, \ex^{c_1})$ and $1- \tau^2< c_3/c_2$.
Finally, the union bound proves that
\[
\P(\Ec_1^{(n)}) \leq 2^{\kappa_{m,n}m} \ex^{-\frac{dc_2^2(\tau^2-1)^2}{16c_3}},
\]
which completes the proof.
$\hfill \Box$
%


\section{Conclusions}\label{sec:conclusion}

In this paper, we have considered the problem of recovering structured signals from underdetermined  linear measurements.  We have used the Komogorov complexity of the quantized signal as a universal measure of complexity to both unify  many of the models explored in the CS literature, and also provide a framework to analyze future structured signal models. We have shown that, if we consider low-complexity signals, then the minimum complexity pursuit (MCP) scheme inspired by Occam's razor recovers the simplest solution of a set of random linear measurements.
In fact, we  have proved that MCP successfully recovers a signal of ``complexity'' $\kappa_n$ at ambient dimension of $n$ from only $3\kappa_n$ random linear measurements. We  have also considered more practical scenarios where the signal is not exactly low complexity but rather is ``close'' to a low complexity signal. We have shown that, even in such cases, the MCP algorithm
provides a good estimate of the signal from much fewer samples than the ambient dimension of the signal.

As mentioned above, Kolmogorov complexity of a sequence is not computable. However, currently we are working on deriving implementable schemes by replacing the Kolmogorov complexity by computable measures such as  minimum description length \cite{Rissanen86}.

\appendix
\subsection{Review of prefix Kolmogorov Complexity} \label{app:kol}
In an effort to formalize the concept of computability of functions, Turing introduced the notion of  {\em Turing machine} \cite{Turing:36}. A Turing machine is a device that has a finite number of states, a memory that is in the form of a tape, and  a head that at each time points to one of the blocks on the tape. The tape consists of  adjacent blocks, each of which can store one of the three symbols $\mathcal{I}= \{0,1, B \}$, where $B$ represents a blank. Initially the code $\sv\in\{0,1\}^*$ is written on the tape in adjacent blocks, and the rest of the tape is filled with blanks. The machine starts from the leftmost non-blank symbol on the tape, and it  works in discrete time steps. At every time instance, it reads the symbol from the tape that  the head is pointing to, and based on its current state and the acquired  information from the tape, it performs the following actions:
\begin{enumerate}
\item update the state,
\item write one symbol from $\mathcal{I}$ onto the tape at the location the head is pointing to,
\item move the head one block  to the right.
\end{enumerate}
The process continues until the machine enters the halting state. The output of a Turing machine $ {\tt T}$  given  is defined as  follows. If the machine does not halt, then $ {\tt T}(\sv)$ is not defined.  If $ {\tt T}$ halts, then the tape contains a binary string that is surrounded by blanks. $ {\tt T}(\sv)$, \ie the output of $ {\tt T}$ given $\sv$, is set to this binary string. If  the output string contains blanks between the binary symbols, then they are replaced by zeros to make the output a binary sequence.
Note that by construction, if both $ {\tt T}(\sv_1)$ and $ {\tt T}(\sv_2)$ are defined, then none of them can be a prefix of the other one.  There are alternative constructions of Turing machines that do not guarantee this property \cite{book_vitanyi}, but in this paper we only consider those that have this property.


One of the most fundamental results in the algorithmic information theory is the existence of { universal machines} that are additively optimal (see Theorem 2.1.1 in \cite{book_vitanyi}). A {\em universal  machine} ${\tt U}$ is a machine that is able to imitate the behavior of all Turing machines on any input string. A universal  machine ${\tt U}$ is (additively) optimal, if for every Turing machine $ {\tt T}$, there exists a constant $c_ {\tt T}$ that only depends on $ {\tt T}$, such that 
\[
\min\{\ell(\sv): {\tt U}(\sv)=\xv\} \leq \min\{\ell(\sv'): {\tt T}(\sv')=\xv\} +c_{ {\tt T}}.
\]
The existence of  optimal universal Turing machines is a result of the fact that any Turing machine can be uniquely specified with a finite number of bits. (Refer to Chapter 1 of \cite{book_vitanyi, CoGaGr89} for more information on the universal Turing Machines.) Given optimal universal machine $ {\tt U}$, the {\em prefix Kolmogorov complexity} of $\xv\in\{0,1\}^{*}$ with respect to $ {\tt U}$ is defined as
\[
 {K}_{{\tt U}} (\x) \triangleq \min\{\ell(\sv): {\tt U}(\sv)=\xv\}.
\]

\subsection{Proof of Theorem \ref{thm:properties}} \label{app:proof_thmprop}

\begin{itemize}
\item[i.] The following program prints $\xv$ : Print the following bit sequence $x_1, x_2, \ldots, x_{\ell(\x)}$. The first part that explains the structure has a constant length, c, and then the bits themselves require $\ell(\x)$ bits. Therefore, the length of the program is less than $\ell(\x)+c$.

\item[ii.] Let $\pb_\x$ and $\pb_{\y}$ denote the shortest programs that print $\x$ and $\y$ respectively. The following program prints $(\x,\y)$:
Print a concatenation of two numbers and the programs for these two numbers are $\pb_\x$ and $\pb_{\y}$. 

Note that since the programs are assumed to be prefix free, after the explanation, ``Print a concatenation of two numbers'', the machines continues until it goes into the halting state. At this point it has already printed $\xv$ . But since it knows that we expect another number, it again starts to read the bits and therefore will print $\y$ as well.

\item[iii.] The proof of this part is also straightforward, since using a constant number of bits,  the code can be required to ignore the extra information $\y$, and then use the code that achieve $K(x)$.

\item[iv.] We use the same program that we used in Part 1. Notice that since the machine does not know $\ell(\x)$ we should spend $K(\ell(\x))$ bits to describe this number as well.  Hence, overall we require $K(\x \, | \, \ell(\x)) + K(\ell(\x))+c$ bits.

\item[v.] First note that the length of the binary representation of $n$ which is denoted by $\ell(n)$ is $\log n$. According to Part iv we have
 \begin{align}
K(n) &\leq K(n \, | \, \ell(n)) + K(\ell(n))+c \leq \ell(n)+ 2 \max(\log(\log n) ,1) + c'  \nonumber \\
         &\leq \log n + 2 \max(\log\log n,1) + c'. 
\end{align}

\item[vi.] The proof is very similar to the proof of Part ii, and hence we skip it.

\end{itemize}

\bibliographystyle{unsrt}
\bibliography{myrefs}

\end{document}